\documentclass[
 preprint,
 amsmath,amssymb,
 aps,
 pre,
]{revtex4-2}

\usepackage{graphicx}
\usepackage{bm}
\usepackage{booktabs}
\usepackage{xcolor}

\begin{document}

\title{Weakly Nonlinear Whitham Modulation and Dispersive Shock Waves in the Modified Gardner Equation}

\author{Nese Ozdemir}
\email{nese.ozdemir@marmara.edu.tr}
\affiliation{Marmara University, Department of Computer Engineering, Faculty of Technology, Istanbul, 34840,  Turkey}

\author{Ali Demirci}
\email{demircial@itu.edu.tr}
\thanks{Corresponding author.}
\author{Ayse Peker Dobie}
\email{pdobie@itu.edu.tr}
\affiliation{Istanbul Technical University, Department of Mathematics Engineering,
Faculty of Science and Letters, Istanbul, 34469, Turkey}

\date{\today}

\begin{abstract}
We investigate weakly nonlinear Whitham modulation and dispersive shock waves (DSWs) in a modified Gardner equation. A three-equation modulation system is derived through a Stokes expansion and averaged conservation laws. Perturbative analysis of its characteristic structure provides explicit conditions for hyperbolicity and genuine nonlinearity, and identifies a wavenumber-dependent transition between real and complex characteristic splitting. The dispersive resolution of compressive Riemann data is examined using DSW fitting, yielding analytical predictions for the harmonic- and solitary-wave edge velocities. The solitary-wave prediction is further supported by a traveling-wave energy analysis. Numerical continuation of the modulation characteristics identifies the limitations of the weakly nonlinear approximation, while direct numerical simulations using a Fourier pseudospectral method confirm close agreement with the predicted leading-edge velocity and the leading solitary-wave amplitude for the resolved cases. These results clarify the weakly nonlinear characteristic structure and the edge dynamics of dispersive shocks for the modified Gardner equation.
\end{abstract}


\maketitle

\section{Introduction}
\label{sec:introduction}

The competition between nonlinear steepening and dispersion is a fundamental mechanism in the evolution of long waves. When a compressive initial profile develops a gradient catastrophe in its dispersionless approximation, a conservative dispersive regularization can replace the resulting discontinuity by an expanding, rapidly oscillating wavetrain, commonly termed a dispersive shock wave (DSW). The classical description of such a transition originates in the work of Gurevich and Pitaevskii on the Korteweg--de Vries (KdV) equation and in Whitham's averaging theory for slowly modulated periodic waves~\cite{GurevichPitaevskii1974,Whitham1965}. The resulting modulation framework connects the oscillatory interior to a small-amplitude, harmonic edge and a long-period, solitary-wave edge. Its integrable KdV formulation was subsequently developed through inverse-spectral and multiphase averaging methods~\cite{FlaschkaForestMcLaughlin1980,El1996}. These classical results provide a benchmark, but they do not by themselves determine the modulation structure of a general nonlinear dispersive conservation law.

Extensions of KdV-type models arise when higher-order nonlinear contributions are retained in long-wave asymptotics. In particular, extended and extended--extended KdV approximations account for nonlinear effects beyond the conventional quadratic KdV truncation ~\cite{MarchantSmyth2006,MartinTseluikoKhusnutdinova2026,BaqerHorikisFrantzeskakis2025}. Norevik and Kalisch~\cite{NorevikKalisch2024} studied the formulation of energy conservation for an extended--extended KdV (eeKdV) model of unidirectional surface waves. Related higher-order and Gardner-type reductions arise in stratified-fluid flows, plasma models, and multicomponent condensates, where the balance between competing nonlinear terms can influence wave profiles and their dispersive evolution~\cite{KamchatnovEtAl2012,KamchatnovEtAl2013,TamangSaha2020,KamchatnovEtAl2014}. The Gardner equation is especially instructive because the additional cubic nonlinearity may change the topology of dispersive Riemann solutions; its undular-bore theory and forced-flow applications have been investigated in detail~\cite{KamchatnovEtAl2012,KamchatnovEtAl2013}. Cylindrical and spherical Gardner models further illustrate how geometric effects alter nonlinear modulation~\cite{AslanovaAhmetolanDemirci2020,AslanovaDemirciAhmetolan2022}. The present work instead considers a scalar, constant-coefficient equation with quadratic, cubic, and quartic nonlinear contributions to the conservation-law flux. The precise physical scaling and the retained asymptotic terms are specified in Sec.~\ref{subsec:physical-origin}; the model is not intended to describe every physical regime covered by the preceding examples.

For nonintegrable dispersive equations, a complete diagonalization or explicit self-similar solution of the finite-amplitude Whitham system is generally unavailable. El~\cite{El2005} showed that the limiting behavior at the harmonic and solitary-wave edges can nevertheless yield DSW transition conditions without solving the entire modulation system. This DSW-fitting viewpoint, and its assumptions and limitations, are discussed systematically by El and Hoefer~\cite{ElHoefer2016}; subsequent developments include generalized nonintegrable DSW constructions~\cite{Kamchatnov2019} and extensions to Benjamin--Ono-type nonlocal dispersion~\cite{ElNguyenSmyth2018}. These results motivate an important distinction: predicting edge velocities from reduced dispersion relations is a less restrictive task than constructing a globally admissible, finite-amplitude modulation solution. In particular, edge matching alone does not establish the shape of the interior oscillatory envelope.

The characteristic type of the modulation equations introduces a separate issue. The existence of a periodic traveling wave does not guarantee that its slowly varying parameters satisfy a hyperbolic modulation system. For generalized KdV equations, the connection between the averaged Whitham equations and the long-wavelength spectral behavior of periodic waves has been investigated by Johnson, Zumbrun, and Bronski and by Johnson and Zumbrun~\cite{JohnsonZumbrunBronski2010,JohnsonZumbrun2010}. More general modulational-instability criteria and the geometry of transitions between real and complex characteristics have also been studied~\cite{BridgesRatliff2017,BridgesRatliff2019,BinswangerHoeferIlan2021,Lighthill1965,JinLiaoLin2019}. Such a transition is significant because real characteristic speeds are a necessary local ingredient in the usual simple-wave DSW construction, whereas complex characteristics signal the failure of that hyperbolic description. However, a leading-order weak-amplitude discriminant must not be interpreted as an exact finite-amplitude spectral-stability boundary.

The broader dispersive-hydrodynamic literature further demonstrates why the type and admissibility of a modulated wavetrain should be assessed rather than presumed. Nonconvex linear dispersion can produce radiating or traveling DSWs in fifth-order KdV-type equations~\cite{SprengerHoefer2017,HoeferSmythSprenger2019}; admissible discontinuities of the modulation system may also arise as zero-dispersion limits of heteroclinic traveling waves~\cite{SprengerHoefer2020}. The Benjamin--Bona--Mahony dispersive Riemann problem exhibits additional regimes, including multiphase wavetrains and dispersive shocks, with a pronounced dependence on the smoothing of the initial step~\cite{CongyElHoeferShearer2021}. These phenomena provide useful comparisons but involve dispersion mechanisms distinct from the third-order local dispersion studied here. They are therefore used only as context, not as direct analogs of the present model. Accordingly, neither the presence of a hyperbolic--elliptic modulation transition nor the existence of nonclassical DSW structures in those models implies that the same wave patterns occur in the present equation.

Motivated by these developments, we investigate the modified Gardner equation
\begin{equation}
 v_{\tau}+f(v)v_{\xi}+\varepsilon^{2}v_{\xi\xi\xi}=0,
 \qquad f(v)=v-v^{2}+2v^{3},\qquad 0<\varepsilon\ll1.
 \label{eq:intro-model}
\end{equation}
The associated dispersionless flux $F$, defined by $F'(v)=f(v)$, is strictly convex because
$f'(v)=1-2v+6v^{2}>0$ for every real $v$, while the curvature $f''(v)=12v-2$ changes sign.
Thus the model allows us to investigate a modulation-type transition without a loss of convexity of the dispersionless flux.

The main contributions of this work are fourfold. First, we derive a weakly nonlinear three-component Whitham system from a Stokes expansion and averaged conservation laws, and determine its leading-order characteristic splitting, hyperbolicity, and genuine nonlinearity. Second, we identify the wavenumber-dependent discriminant
\begin{equation}
 \Delta(m,k)=[f'(m)]^{2}-3k^{2}f''(m),
 \label{eq:intro-discriminant}
\end{equation}
which distinguishes real from complex splitting of the two coalescing modulation characteristics away from the transition curve. Third, we study compressive Riemann data by DSW fitting, derive the two edge velocities, and relate the solitary-wave edge to the exact amplitude--speed relation obtained from the traveling-wave energy integral. Fourth, we use Fourier pseudospectral simulations to assess the leading-edge propagation speed and to compare the predicted solitary-wave amplitudes and crest-aligned profiles with direct solutions for several compressive jumps.

The scope of this work is limited to weakly nonlinear modulation and edge fitting. We do not construct a global finite-amplitude Whitham solution or a fully validated analytic DSW envelope; these limitations are discussed in Sec.~\ref{subsec:edge-hyperbolicity}.

The remainder of the paper is organized as follows. Sec.~\ref{sec:whitham} establishes the physical scaling, traveling-wave formulation, and weakly nonlinear Whitham characteristic structure. Sec.~\ref{sec:riemann_dsw} develops the dispersive Riemann problem and its harmonic- and solitary-wave edge conditions. Sec.~\ref{sec:numerical} presents numerical diagnostics and direct comparisons with the dispersive PDE. The conclusions summarize the scope of the results and discuss the unresolved problems of finite-amplitude modulation and the hyperbolic--elliptic transition.

\section{Weakly Nonlinear Whitham Modulation Theory}
\label{sec:whitham}
We begin by establishing the physical origin and nondimensional
scaling of the governing equation, with particular attention to
the asymptotic assumptions underlying its derivation. We then
examine the periodic traveling-wave problem and construct a
weakly nonlinear Stokes expansion, which provides the quantities
required for averaging the conservation laws. The resulting
Whitham modulation system is subsequently analyzed through its
characteristic velocities and eigenvectors, allowing us to
identify the conditions for strict hyperbolicity, the onset of
ellipticity, and genuine nonlinearity. This framework provides
the theoretical basis for the dispersive Riemann problem and
DSW-edge analysis developed in the following section.

\subsection{Physical origin and governing equation}
\label{subsec:physical-origin}

The equation studied here is motivated by the extended--extended
Korteweg--de Vries (eeKdV) approximation for unidirectional surface
gravity waves over a horizontal bottom. In the dimensional formulation
of Norevik and Kalisch~\cite{NorevikKalisch2024}, the free-surface
displacement $\eta(x,t)$ relative to the undisturbed water depth $h_0$
satisfies
\begin{equation}
\eta_t+c_w\eta_x
+\frac{3c_w}{2h_0}\eta\eta_x
-\frac{3c_w}{8h_0^2}\eta^2\eta_x
+\frac{3c_w}{16h_0^3}\eta^3\eta_x
+\frac{c_wh_0^2}{6}\eta_{xxx}=0,
\qquad c_w=\sqrt{gh_0},
\label{eq:dimensional-eekdv}
\end{equation}
where $g$ denotes gravitational acceleration. This formal asymptotic
model assumes an inviscid, incompressible and irrotational fluid,
negligible surface tension and transverse effects, and predominantly
right-propagating long waves.Extended KdV models that retain higher-order nonlinear and dispersive corrections have been derived and analyzed for surface and internal waves, including their undular bores, resonant dispersive shock waves, and solitary-wave dynamics~\cite{MarchantSmyth2006,BaqerSmyth2023,BaqerHorikisFrantzeskakis2025,MartinTseluikoKhusnutdinova2026}; Eq.~\eqref{eq:dimensional-eekdv} retains only the nonlinear corrections. If $a_0$ and $\Lambda$ are characteristic
surface-displacement and horizontal-length scales, respectively, the
relevant nondimensional parameters are
\begin{equation}
\alpha=\frac{a_0}{h_0},\qquad
\beta=\frac{h_0^2}{\Lambda^2},\qquad
0<\alpha\ll1,\quad 0<\beta\ll1.
\label{eq:physical-small-parameters}
\end{equation}
The classical KdV balance uses $\beta=O(\alpha)$, whereas the eeKdV
approximation retains the quadratic and cubic nonlinear corrections
under the distinguished ordering $\beta=O(\alpha^3)$. At the level of
the original asymptotic expansion, the omitted contributions are of
orders $O(\alpha^4,\alpha\beta,\beta^2)$; thus
\eqref{eq:dimensional-eekdv} is understood as a truncated model,
not as an exact representation of the full Euler free-surface problem.

To relate this physical formulation to the mathematical normalization
used in the modulation analysis, introduce an arbitrary fixed
horizontal scale $L>0$ and set
\begin{equation}
\xi=\frac{x-c_wt}{L},\qquad
\tau=\frac{6c_wt}{L},\qquad
\eta(x,t)=4h_0v(\xi,\tau),\qquad
\varepsilon=\frac{h_0}{6L}.
\label{eq:physical-normalization}
\end{equation}
The moving coordinate removes the linear transport at the long-wave
speed $c_w$. Substituting \eqref{eq:physical-normalization} into
\eqref{eq:dimensional-eekdv} and dividing by $24c_wh_0/L$
gives the dimensionless modified Gardner equation
\begin{equation}
v_{\tau}
+\varepsilon^2 v_{\xi\xi\xi}
+v v_{\xi}
-v^2v_{\xi}
+2v^3v_{\xi}=0,
\label{eq:mg}
\end{equation}
where $0<\varepsilon\ll1$ corresponds to $L\gg h_0$.
The transformation is an exact change of variables for the truncated
eeKdV equation, but it does not remove the physical restrictions under
which that equation was originally derived. In particular,
$|\eta|/h_0=4|v|\ll1$ is needed for a direct weak-amplitude
free-surface interpretation. The finite $O(1)$ normalized states
considered later in the Riemann problems are therefore used to
investigate the mathematical dynamics of the reduced dispersive model;
they should not, without additional justification, be interpreted as
quantitatively valid small-amplitude water-wave states. Moreover,
$\varepsilon\ll1$ by itself does not enforce the distinguished
physical ordering $\beta=O(\alpha^3)$.

We introduce the flux function and its primitive
\begin{equation}
f(v)=v-v^2+2v^3,
\qquad
F(v)=\frac{1}{2}v^2-\frac{1}{3}v^3+\frac{1}{2}v^4,
\label{eq:fF}
\end{equation}
so that $F'(v)=f(v)$ and \eqref{eq:mg} takes the conservative form
\begin{equation}
v_{\tau}
+\partial_{\xi}
\left(
F(v)+\varepsilon^2 v_{\xi\xi}
\right)=0.
\label{eq:mg-conservative}
\end{equation}
Equation~\eqref{eq:mg} belongs to the class of generalized KdV (gKdV) equations with a polynomial flux. For this class, the relation between the Whitham modulation equations and the long-wavelength spectral stability of periodic waves has been established rigorously~\cite{JohnsonZumbrunBronski2010,JohnsonZumbrun2010}, nonlinear modulational instability has been proved~\cite{JinLiaoLin2019}, and the propagation of solitons along large-scale background waves has been described~\cite{KamchatnovShaykin2023}. Gardner-type equations with competing quadratic and cubic nonlinearities, including forced, cylindrical, and spherical variants and a modified Gardner equation for plasma waves, provide closely related examples~\cite{KamchatnovEtAl2012,KamchatnovEtAl2013,AslanovaAhmetolanDemirci2020,AslanovaDemirciAhmetolan2022,TamangSaha2020}.

With the governing equation and its flux function established, we now turn to the construction of weakly nonlinear periodic wavetrains.

\subsection{Stokes expansion and the averaged Whitham system}
\label{subsec:stokes-whitham}

We now construct a Stokes expansion of periodic traveling waves and use it to average the conservation laws of the governing equation. The procedure follows Whitham's averaged-conservation-law approach~\cite{Whitham1965,Whitham1974}; see Refs.~\cite{Kamchatnov2000,ElHoefer2016} for its use in the theory of dispersive shock waves. Whitham modulation equations of nonintegrable models and their small-amplitude limits have recently been used to determine modulational instability and convexity properties~\cite{BinswangerHoeferIlan2021,HoeferVainchtein2026}. The weakly nonlinear modulation theory developed below describes slowly
varying periodic wavetrains in terms of their mean $m$, amplitude $a$, and
wavenumber $k$. These quantities are allowed to evolve on the slow
$(\xi,\tau)$ scales, while the rapid oscillations are represented by a
phase variable $\theta$.

Let the rapid phase $\theta$ satisfy
\begin{equation}
\theta_{\xi}=\frac{k(\xi,\tau)}{\varepsilon},
\qquad
\theta_{\tau}=-\frac{\omega(\xi,\tau)}{\varepsilon},
\label{eq:phase}
\end{equation}
where $k$ and $\omega$ denote the local wavenumber and frequency,
respectively. The compatibility condition
$\theta_{\xi\tau}=\theta_{\tau\xi}$ immediately yields the conservation
of waves,
\begin{equation}
k_{\tau}+\omega_{\xi}=0.
\label{eq:wave-conservation}
\end{equation}

We seek a multiple-scale expansion of the form
\begin{equation}
v(\theta,\xi,\tau)
=
v_0(\theta,\xi,\tau)
+\varepsilon v_1(\theta,\xi,\tau)
+\cdots ,
\label{eq:multiple-scale}
\end{equation}
where each $v_j$ is $2\pi$-periodic in $\theta$. At leading order,
substitution of \eqref{eq:phase}--\eqref{eq:multiple-scale} into
\eqref{eq:mg} gives
\begin{equation}
-\omega v_{0,\theta}
+k^3v_{0,\theta\theta\theta}
+k f(v_0)v_{0,\theta}=0.
\label{eq:leading-profile}
\end{equation}
Introducing the local phase velocity $c=\omega /k$ and integrating once with respect to $\theta$, the periodic traveling-wave
equation becomes
\begin{equation}
k^2v_{0,\theta\theta}
-cv_0+F(v_0)=C_1,
\label{eq:profile-integrated}
\end{equation}
where $C_1$ is an integration constant.

Before constructing the small-amplitude expansion, it is useful to
separate the existence of periodic traveling waves from their stability
under slowly varying perturbations. Multiplying
\eqref{eq:profile-integrated} by $v_{0,\theta}$ and integrating yields
\begin{equation}
\frac{k^2}{2}v_{0,\theta}^2+\mathcal V(v_0;c,C_1)=E,
\qquad
\mathcal V(v;c,C_1)=\int^v F(s)\,\mathrm{d}s-\frac{c}{2}v^2-C_1v,
\label{eq:profile-energy}
\end{equation}
where $E$ is constant along the orbit. A nonconstant periodic profile
oscillates between two adjacent simple turning points $v_-<v_+$
with $E-\mathcal V(v)>0$ for $v\in(v_-,v_+)$. Its phase period is
\begin{equation}
\mathcal T_\theta
=\sqrt{2}\,k\int_{v_-}^{v_+}
\frac{\mathrm{d}v}{\sqrt{E-\mathcal V(v;c,C_1)}}.
\label{eq:profile-period}
\end{equation}
For the $2\pi$-periodic phase convention used here, the parameters
must satisfy $\mathcal T_\theta=2\pi$. In particular, a constant
state $v_0=m$ is a center of the traveling-wave phase plane when
$\mathcal V''(m)=f(m)-c>0$. In the small-amplitude limit, the period
condition gives $f(m)-c=k^2>0$, consistently with the linear
phase-speed relation derived below. This phase-plane condition
establishes local periodic-orbit existence; it is not a dynamical
stability criterion for the underlying evolution equation. For the KdV equation these periodic orbits are the classical cnoidal waves~\cite{BenjaminLighthill1954}; for the modified KdV equation, the modulational stability of cnoidal waves is determined by the root structure of the corresponding oscillator polynomial~\cite{DriscollONeil1976}.

The stability relevant to dispersive-shock modulation concerns
long-wavelength perturbations of these periodic wavetrains. It is
assessed below through the characteristic speeds of the averaged
Whitham equations, rather than by a full Floquet--Bloch spectral
analysis. Specifically, for sufficiently small nonzero amplitude,
the leading-order splitting of the two coincident zero-amplitude
Whitham characteristics is governed by the discriminant
$\Delta(m,k)$ introduced in \eqref{eq:intro-discriminant}.
A positive $\Delta$ gives real distinct leading-order characteristic
branches, while a negative $\Delta$ signals an elliptic
small-amplitude modulation regime. The boundary $\Delta=0$ requires
higher-order analysis. Thus, the hyperbolicity conditions developed
later provide a local, long-wave modulational diagnostic and should
not be interpreted as a proof of spectral stability against arbitrary
perturbation wavelengths.

For weakly nonlinear periodic waves, we introduce the Stokes expansion
\begin{equation}
v_0
=
m+a\cos\theta
+c_4a^2\cos(2\theta)
+O(a^3),
\label{eq:stokes-wave}
\end{equation}
together with
\begin{equation}
c=c_0+a^2c_2+O(a^4).
\label{eq:stokes-speed}
\end{equation}
Here $m=m(\xi,\tau)$ is the local mean, $a=a(\xi,\tau)$ is the
fundamental harmonic amplitude, and $k=k(\xi,\tau)$ is the local
wavenumber.

Substituting \eqref{eq:stokes-wave}--\eqref{eq:stokes-speed} into
\eqref{eq:profile-integrated} and balancing harmonics at successive orders
gives
\begin{equation}
c_0=f(m)-k^2,
\qquad
c_4=\frac{f'(m)}{12k^2},
\qquad
c_2=\frac{[f'(m)]^2}{24k^2}+\frac{f''(m)}{8}.
\label{eq:c0-c4-c2}
\end{equation}
For the modified Gardner nonlinearity, \(f'(m)=1-2m+6m^2\) and \(f''(m)=12m-2\), so that \(c_4=(1-2m+6m^2)/(12k^2)\) and \(c_2=[(1-2m+6m^2)^2+6k^2(6m-1)]/(24k^2)\).
Consequently, the weakly nonlinear dispersion relation can be
written as
\begin{equation}
\omega(m,a,k)
=
k\left[
f(m)-k^2+a^2c_2(m,k)
\right]
+O(a^4).
\label{eq:nonlinear-dispersion}
\end{equation}
Its zero-amplitude limit,
\begin{equation}
\omega_0(m,k)=kf(m)-k^3,
\label{eq:linear-dispersion}
\end{equation}
is the linear dispersion relation for small-amplitude waves propagating
on the slowly varying mean state $m$.

The modulation equations are obtained by averaging independent
conservation laws of \eqref{eq:mg} over one period of the rapid phase \cite{Whitham1965,Whitham1974}.
For a $2\pi$-periodic function $g(\theta)$, we use the notation
\begin{equation}
\langle g\rangle
=
\frac{1}{2\pi}
\int_0^{2\pi}g(\theta)\,\mathrm{d}\theta.
\label{eq:average}
\end{equation}

Averaging the mass conservation law \eqref{eq:mg-conservative} and using
the Stokes approximation \eqref{eq:stokes-wave} gives, through
$O(a^2)$,
\begin{equation}
m_{\tau}
+
\partial_{\xi}
\left[
F(m)+\frac{a^2}{4}f'(m)
\right]
=0.
\label{eq:whitham-conservation-1}
\end{equation}

A second conservation law follows by multiplying \eqref{eq:mg} by $v$.
It can be written as
\begin{equation}
\partial_{\tau}\left(\frac{v^2}{2}\right)
+
\partial_{\xi}
\left[
H(v)
+\varepsilon^2
\left(
vv_{\xi\xi}-\frac{1}{2}v_{\xi}^2
\right)
\right]=0,
\label{eq:second-conservation}
\end{equation}
where
\begin{equation}
H'(v)=vf(v),
\qquad
H(v)
=
\frac{1}{3}v^3-\frac{1}{4}v^4+\frac{2}{5}v^5.
\label{eq:H}
\end{equation}
Period averaging yields
\begin{equation}
\left(
\frac{m^2}{2}+\frac{a^2}{4}
\right)_{\tau}
+
\partial_{\xi}
\left\{
H(m)
+
\frac{a^2}{4}
\left[
f(m)+mf'(m)-3k^2
\right]
\right\}
=0.
\label{eq:whitham-conservation-2}
\end{equation}

Equations \eqref{eq:whitham-conservation-1} and
\eqref{eq:whitham-conservation-2}, together with the conservation of
waves
\begin{equation}
k_{\tau}
+
\partial_{\xi}
\left\{
k\left[
f(m)-k^2+a^2c_2(m,k)
\right]
\right\}
=0,
\label{eq:whitham-conservation-3}
\end{equation}
constitute the weakly nonlinear Whitham modulation system in conservative
form, with modulation variables \(\mathbf q=(m,a,k)^{T}\).

For the characteristic analysis, it is useful to recast the averaged
system in quasilinear form,
\begin{equation}
\mathbf q_{\tau}
+
A(\mathbf q)\mathbf q_{\xi}=0.
\label{eq:whitham-quasilinear}
\end{equation}
Retaining all contributions consistent with the $O(a^2)$ Stokes
approximation gives
\begin{equation}
A=
\begin{pmatrix}
f+\dfrac{a^2}{4}f''
&
\dfrac{a}{2}f'
&
0
\\[3mm]
af'
&
f-3k^2
&
-3ak
\\[3mm]
kf'
+a^2\left(
\dfrac{f'f''}{12k}+\dfrac{3k}{2}
\right)
&
\dfrac{a}{12k}
\left(
(f')^2+3k^2f''
\right)
&
f-3k^2
+a^2\left(
\dfrac{f''}{8}
-\dfrac{(f')^2}{24k^2}
\right)
\end{pmatrix}
+O(a^3),
\label{eq:whitham-matrix}
\end{equation}
where $f$, $f'$, and $f''$ are evaluated at $m$.

The conservative representation
\eqref{eq:whitham-conservation-1}--\eqref{eq:whitham-conservation-3}
and the quasilinear representation \eqref{eq:whitham-quasilinear}
serve distinct purposes. The former retains the averaged balance
structure and will be used when discontinuous modulation states are
considered, whereas the latter provides the natural framework for the
analysis of characteristic velocities, hyperbolicity, and genuinely
nonlinear characteristic fields.

\subsection{Characteristic structure, hyperbolicity, and genuine nonlinearity}
\label{subsec:characteristic-hyperbolicity}

Having obtained the averaged modulation system, we now examine its characteristic structure, hyperbolicity, and genuine nonlinearity. Since the modulation
matrix is obtained as an asymptotic expansion in the wave amplitude,
the characteristic velocities are sought consistently in powers of
$a$. At zero amplitude, the coefficient matrix reduces to
\begin{equation}
A_0=
\begin{pmatrix}
f(m) & 0 & 0\\
0 & f(m)-3k^2 & 0\\
kf'(m) & 0 & f(m)-3k^2
\end{pmatrix}.
\label{eq:A0}
\end{equation}
Its characteristic polynomial is
\begin{equation}
\det(A_0-\lambda I)
=
\bigl(f(m)-\lambda\bigr)
\bigl(f(m)-3k^2-\lambda\bigr)^2.
\label{eq:A0-charpoly}
\end{equation}
Thus, the zero-amplitude characteristic velocities consist of the
simple branch and the double branch,
\begin{equation}
\lambda_1^{(0)}=f(m),
\qquad
\lambda_2^{(0)}=\lambda_3^{(0)}=f(m)-3k^2.
\label{eq:lambda-zero}
\end{equation}
The degeneracy of the latter pair at $a=0$ requires a degenerate
perturbation analysis when the finite-amplitude corrections are
introduced.

For the simple characteristic branch, the perturbation calculation
gives
\begin{equation}
\lambda_1
=
f(m)+\frac{a^2}{4}f''(m)+O(a^3).
\label{eq:lambda1-general}
\end{equation}
Using \(f'(m)=1-2m+6m^2\) and \(f''(m)=12m-2\), this becomes
\begin{equation}
\lambda_1
=
m-m^2+2m^3
+a^2\left(3m-\frac{1}{2}\right)
+O(a^3).
\label{eq:lambda1}
\end{equation}

To resolve the double characteristic branch, let
\begin{equation}
\lambda_{\pm}
=
f(m)-3k^2
+a s_{\pm}
+a^2t_{\pm}
+O(a^3).
\label{eq:lambda-pm-expansion}
\end{equation}
The first nontrivial solvability condition obtained from
$\det(A-\lambda I)=0$ is
\begin{equation}
4s_{\pm}^2
-
[f'(m)]^2
+
3k^2f''(m)
=0.
\label{eq:s-solvability}
\end{equation}
Using the discriminant \eqref{eq:intro-discriminant}, we obtain
\begin{equation}
s_{\pm}=\pm\frac{1}{2}\sqrt{\Delta(m,k)},
\qquad
t_{\pm}=-\frac{\Delta(m,k)}{48k^2}.
\label{eq:s-t-pm}
\end{equation}
The two characteristic velocities bifurcating from the double
zero-amplitude branch are therefore
\begin{equation}
\lambda_{\pm}
=
f(m)-3k^2
\pm\frac{a}{2}\sqrt{\Delta(m,k)}
-\frac{a^2}{48k^2}\Delta(m,k)
+O(a^3).
\label{eq:lambda-pm}
\end{equation}
We identify \(\lambda_2=\lambda_-\) and \(\lambda_3=\lambda_+\).

For the modified Gardner nonlinearity, the discriminant assumes the
explicit form
\begin{equation}
\Delta(m,k)
=
(1-2m+6m^2)^2
+6k^2(1-6m).
\label{eq:Delta}
\end{equation}
Equation \eqref{eq:Delta} provides the principal criterion governing
the local characteristic type of the weakly nonlinear modulation
system. In particular, $\lambda_2$ and $\lambda_3$ are distinct and
real whenever $\Delta>0$, coalesce when $\Delta=0$, and form a
complex-conjugate pair when $\Delta<0$.

The characteristic geometry can also be described in terms of the
corresponding right eigenvectors. For the first characteristic family,
a convenient leading-order normalization gives
\begin{equation}
\mathbf r_1
=
\begin{pmatrix}
1\\[1mm]
0\\[1mm]
\dfrac{f'(m)}{3k}
\end{pmatrix}
+O(a).
\label{eq:r1}
\end{equation}
For the two characteristic families emerging from the double
eigenvalue, it is useful to normalize the amplitude component to
unity. With $\sigma=\pm1$, where
$\lambda_{\sigma}=\lambda_{\pm}$, one obtains
\begin{equation}
\mathbf r_{\sigma}
=
\begin{pmatrix}
-\dfrac{a f'(m)}{6k^2}\\[3mm]
1\\[2mm]
-\sigma\dfrac{\sqrt{\Delta(m,k)}}{6k}
\end{pmatrix}
+O(a).
\label{eq:r-pm}
\end{equation}
These eigenvectors determine the local characteristic directions in
the $(m,a,k)$ modulation space. No assumption concerning the existence
of a global set of Riemann invariants is required for the subsequent
analysis.

The characteristic discriminant \eqref{eq:Delta} yields a direct
classification of the weakly nonlinear Whitham system. Since \(f'(m)=1-2m+6m^2>0\) for all real \(m\), as noted in Sec.~\ref{sec:introduction}, the sign change in \(\Delta\) is entirely associated with the curvature \(f''(m)=12m-2\) and the wavenumber-dependent contribution to the characteristic splitting.

For $m<1/6$, one has $f''(m)<0$, and hence
\begin{equation}
\Delta(m,k)
=
[f'(m)]^2+3k^2|f''(m)|>0
\qquad
(k>0).
\label{eq:Delta-left}
\end{equation}
At \(m=1/6\), the wavenumber-dependent contribution vanishes, giving
\(\Delta(1/6,k)=25/36>0\). Thus, no loss of hyperbolicity occurs for
$m\leq1/6$ within the present weakly nonlinear description.

For $m>1/6$, however, $f''(m)>0$ and the discriminant decreases with
increasing wavenumber. The boundary $\Delta=0$ is given explicitly by
\begin{equation}
k=k_c(m)
=
\frac{1-2m+6m^2}
{\sqrt{6(6m-1)}},
\qquad
m>\frac{1}{6}.
\label{eq:kc}
\end{equation}
Consequently, for \(m>1/6\), one has \(\Delta(m,k)>0\) for \(0<k<k_c(m)\)
and \(\Delta(m,k)<0\) for \(k>k_c(m)\). The former case corresponds to real
characteristic velocities and hence to a hyperbolic modulation system,
while the latter is characterized by a complex-conjugate pair of
modulation characteristics. In the Whitham description, this transition
identifies the small-amplitude hyperbolicity boundary of the underlying
periodic wavetrain.

The condition $\Delta<0$ is the counterpart, for the present Whitham system, of the Lighthill criterion for the modulational instability of weakly nonlinear wavetrains~\cite{Lighthill1965,Whitham1974}. For the Gardner equation, the corresponding nonlinear Schr\"odinger reduction shows that wave packets are unstable only for a positive coefficient of the cubic nonlinear term and a sufficiently high carrier wavenumber~\cite{GrimshawEtAl2001}; the role of $f''(m)$ and the threshold $k_c(m)$ in \eqref{eq:kc} are consistent with that result. The geometry of hyperbolic--elliptic transitions in Whitham modulation theory and its connection with the sign characteristic of the modulation equations have been analyzed in Refs.~\cite{BridgesRatliff2017,BridgesRatliff2019}.

Strict hyperbolicity alone does not determine the nonlinear geometry
of the characteristic fields. For this purpose, we define
\begin{equation}
\mu_j=\nabla\lambda_j\cdot\mathbf r_j,
\label{eq:mu-definition}
\end{equation}
where the gradient is taken with respect to the modulation variables
\(\mathbf q=(m,a,k)^{T}\). A characteristic field is genuinely nonlinear
whenever $\mu_j\neq0$ and becomes linearly degenerate on parameter sets
where $\mu_j=0$. Loss of genuine nonlinearity of the modulation system generates nonclassical wave structures, such as kinks and trigonometric bores in the Gardner equation~\cite{KamchatnovEtAl2012}, and modifies soliton--mean-flow interaction~\cite{vanDerSandeElHoefer2021}; see also Refs.~\cite{ElHoefer2016,HoeferVainchtein2026}.

For the first characteristic family,
\eqref{eq:lambda1-general} and \eqref{eq:r1} give
\begin{equation}
\mu_1
=
f'(m)+O(a^2).
\label{eq:mu1}
\end{equation}
Since $f'(m)>0$ globally, the first characteristic field remains
genuinely nonlinear throughout the weakly nonlinear regime.

For the remaining two characteristic families, the degenerate
perturbation structure must again be retained. Using
\eqref{eq:lambda-pm} and the corresponding eigenvectors gives
\begin{equation}
\mu_{\sigma}
=
\sigma\frac{3}{2}\sqrt{\Delta}
+
a
\left[
\frac{[f'(m)]^2}{12k^2}
+\frac{f''(m)}{2}
\right]
+O(a^2),
\qquad
\sigma=\pm1.
\label{eq:mu-pm-general}
\end{equation}
Accordingly,
\begin{align}
\mu_2
&=
-\frac{3}{2}\sqrt{\Delta}
+aB(m,k)
+O(a^2),
\label{eq:mu2}
\\
\mu_3
&=
+\frac{3}{2}\sqrt{\Delta}
+aB(m,k)
+O(a^2),
\label{eq:mu3}
\end{align}
where
\begin{equation}
B(m,k)
=
\frac{(1-2m+6m^2)^2}{12k^2}
+6m-1.
\label{eq:B}
\end{equation}

In the zero-amplitude limit, the loss of genuine nonlinearity of the
second and third fields coincides with the coalescence condition
$\Delta=0$. At finite but small amplitude, however, the $O(a)$
correction in \eqref{eq:mu2}--\eqref{eq:mu3} permits the formal
linear-degeneracy condition to separate from the hyperbolicity
boundary. In particular, for \(m\geq1/6\), one has \(B(m,k)>0\), and
therefore \(\mu_3>0\) throughout the strictly hyperbolic region to the
order considered. The second characteristic field, on the other hand,
formally admits the degeneracy condition
\begin{equation}
-\frac{3}{2}\sqrt{\Delta(m,k)}
+aB(m,k)=0.
\label{eq:mu2-zero}
\end{equation}

The weakly nonlinear expansion imposes an additional restriction on
the interpretation of \eqref{eq:mu2-zero}. Indeed, the second harmonic
in \eqref{eq:stokes-wave} has magnitude
$c_4a^2$, so that its size relative to the fundamental harmonic is
measured by
\begin{equation}
\chi(m,a,k)
=
\left|
\frac{a f'(m)}{12k^2}
\right|.
\label{eq:eta}
\end{equation}
The condition $\chi\ll1$ is required for a well-separated Stokes
hierarchy. In particular, the limit $k\rightarrow0$ is nonuniform
within the present expansion and cannot be interpreted using the same
fixed-amplitude asymptotics.

Figure~\ref{fig:characteristic-structure} summarizes the characteristic
structure for three representative weak amplitudes. The
$\Delta=0$ curve separates the hyperbolic and elliptic regions, while
the line $m=1/6$ marks the change in sign of $f''(m)$. The formal
$\mu_2=0$ locus is shown together with the diagnostic contour
$\chi=0.2$. Over the displayed parameter range, the predicted loss of
genuine nonlinearity of the second characteristic family occurs in a
region where the Stokes hierarchy is already insufficiently
separated according to this conservative diagnostic. Thus, the
$\mu_2=0$ curve should be regarded as a formal prediction of the
truncated weakly nonlinear modulation system rather than as a robust
finite-amplitude degeneracy boundary. No corresponding $\mu_3=0$
locus occurs in the displayed hyperbolic region.

Within a strictly hyperbolic and genuinely nonlinear region, the
characteristic eigenvectors define wave curves in modulation space.
If $s$ denotes the self-similar characteristic parameter with
$s=\lambda_j$, the $j$th characteristic wave curve satisfies
\begin{equation}
\frac{d\mathbf q}{ds}
=
\frac{\mathbf r_j(\mathbf q)}
{\mu_j(\mathbf q)}.
\label{eq:wave-curve-general}
\end{equation}
For the two characteristic families emerging from the double
zero-amplitude eigenvalue, the amplitude itself provides a convenient
local parameter. Using \eqref{eq:r-pm}, the leading-order wave-curve
equations reduce to
\begin{equation}
\frac{dm}{da}
=
-\frac{a f'(m)}{6k^2},
\qquad
\frac{dk}{da}
=
-\sigma
\frac{\sqrt{\Delta(m,k)}}{6k},
\qquad
\sigma=\pm1.
\label{eq:wave-curves-pm}
\end{equation}
For a wave curve issuing from
\begin{equation}
(m,a,k)=(m_0,0,k_0),
\qquad
\Delta(m_0,k_0)>0,
\end{equation}
the local expansions are
\begin{align}
m(a)
&=
m_0
-\frac{f'(m_0)}{12k_0^2}a^2
+O(a^3),
\label{eq:m-wavecurve-local}
\\
k_{\sigma}(a)
&=
k_0
-\sigma
\frac{\sqrt{\Delta(m_0,k_0)}}{6k_0}a
+O(a^2).
\label{eq:k-wavecurve-local}
\end{align}
Thus, the two characteristic families have the same quadratic
variation of the mean at their point of emergence but separate
linearly in wavenumber. For $m_0>1/6$, the branch associated with
$\sigma=-1$ initially moves toward increasing $k$ and therefore toward
the hyperbolicity boundary, whereas the $\sigma=+1$ branch initially
moves toward decreasing $k$ and into the hyperbolic region.

The first characteristic family admits an especially simple
leading-order description. From \eqref{eq:r1}, one has \(dk/dm=f'(m)/(3k)\) and \(da/dm=0\), and hence
\begin{equation}
k^2-\frac{2}{3}f(m)=\mathrm{const}.
\label{eq:first-wavecurve}
\end{equation}
Equations \eqref{eq:wave-curves-pm} and
\eqref{eq:first-wavecurve} provide a characteristic description of
the weakly nonlinear modulation dynamics without requiring the
existence of a global Riemann-invariant coordinate system. They also
provide the natural wave-curve framework for the self-similar
modulation problem considered in the following section.

\section{The Riemann Problem and Dispersive Shock Fitting}
\label{sec:riemann_dsw}

We now investigate the dispersive resolution of a compressive
Riemann problem for the modified Gardner equation. The
characteristic analysis of Sec.~\ref{sec:whitham} provides a local
description of weakly nonlinear periodic wavetrains,
including their hyperbolicity properties and characteristic
geometry. The construction of a DSW, however, requires additional information concerning the matching of the modulated oscillatory region to the
nonoscillatory states on either side.

For integrable equations, this matching can be formulated
within an exact Whitham system expressed in suitable Riemann
invariants ~\cite{GurevichPitaevskii1974,FlaschkaForestMcLaughlin1980,KamchatnovEtAl2012}. In the present setting, no such exact diagonal
representation is assumed. Instead, we employ the
DSW-fitting approach ~\cite{El2005,ElHoefer2016}, which extracts the harmonic- and
solitary-wave edge information from the dispersionless
characteristic velocity and the linear and conjugate
dispersion relations. The resulting edge predictions are
then examined in relation to the characteristic structure
of the weakly nonlinear modulation system. The fitting method has also been applied to nonlocal, higher-order, and nonconvex-dispersion models and to the Benjamin--Bona--Mahony equation~\cite{ElNguyenSmyth2018,SprengerHoefer2017,BaqerHorikisFrantzeskakis2025,CongyElHoeferShearer2021}.

\subsection{Dispersionless dynamics and the Riemann problem}
\label{subsec:dispersionless}

Consider the step initial condition
\begin{equation}
v(\xi,0)=
\begin{cases}
v_-, & \xi<0,\\
v_+, & \xi>0,
\end{cases}
\qquad v_->v_+.
\label{eq:riemann_data}
\end{equation}
The two constant states are taken to be independent of
the dispersion parameter $\varepsilon$.

In the formal dispersionless limit, the governing equation
reduces to the scalar conservation law
\begin{equation}
v_\tau+F(v)_\xi=0,
\label{eq:dispersionless}
\end{equation}
where
\begin{equation}
F(v)=\frac12v^2-\frac13v^3+\frac12v^4,
\qquad
F'(v)=f(v)=v-v^2+2v^3.
\label{eq:riemann_flux}
\end{equation}
The corresponding characteristic velocity is $f(v)$.
Since \(F''(v)=f'(v)=1-2v+6v^2=6(v-1/6)^2+5/6>0\) for every \(v\in\mathbb{R}\), the dispersionless flux is strictly convex.

Consequently, the decreasing step in
\eqref{eq:riemann_data} produces an entropy-admissible
compressive shock in the dispersionless problem. Its
propagation speed is determined by the Rankine--Hugoniot
condition,
\begin{equation}
s_{\mathrm{RH}}
=\frac{F(v_-)-F(v_+)}{v_--v_+}
=\frac{1}{v_--v_+}\int_{v_+}^{v_-}f(u)\,\mathrm{d}u.
\label{eq:rh_speed}
\end{equation}
Strict convexity implies the Lax inequalities \(f(v_+)<s_{\mathrm{RH}}<f(v_-)\).

For small nonzero dispersion, a compressive jump may be
resolved by an expanding oscillatory region. We adopt the
single-phase, KdV-type DSW-fitting ansatz appropriate to the
positive third-order dispersive term: a small-amplitude
harmonic edge adjacent to the left state $v_-$ and a
solitary-wave edge adjacent to the right state $v_+$.
The existence of a globally matched single-phase DSW is an
assumption of this fitting construction, rather than a
consequence of the strict convexity of $F$. Its local
consistency and its limitations are examined below.

We denote their trajectories by
$\xi_h(\tau)$ and $\xi_s(\tau)$, respectively.
In a self-similar DSW-fitting description, the leading edge velocities are constants, \(\xi_h(\tau)\sim s_h\tau\) and \(\xi_s(\tau)\sim s_s\tau\), where \(s_h\) and \(s_s\) are determined below.

The Rankine--Hugoniot velocity provides a useful
dispersionless reference, but it is not identified with
either DSW edge velocity.

\subsection{DSW edge fitting: harmonic- and solitary-wave edges}
\label{subsec:edge-fitting}

At the harmonic edge, the oscillation amplitude tends
to zero while the local wavenumber approaches a finite
value. Accordingly, the nonlinear dispersion relation
derived in Sec.~\ref{sec:whitham} reduces to
\begin{equation}
\omega_0(m,k)=k f(m)-k^3.
\label{eq:harmonic_dispersion}
\end{equation}
The corresponding group velocity is \(\partial_k\omega_0=f(m)-3k^2\).

The harmonic-edge reduction of the modulation equations
is obtained by combining the dispersionless mean equation
with conservation of waves ~\cite{El2005,ElHoefer2016}:
\begin{equation}
m_\tau+f(m)m_\xi=0,
\qquad
k_\tau+\partial_\xi\omega_0(m,k)=0.
\label{eq:harmonic_mean}
\end{equation}
The characteristic compatibility relation associated
with the dispersionless branch is
\begin{equation}
\frac{dk}{dm}
=
\frac{\partial_m\omega_0}
{f(m)-\partial_k\omega_0}.
\label{eq:harmonic_fitting_ode}
\end{equation}
Substitution of \eqref{eq:harmonic_dispersion} gives
$dk/dm=f'(m)/(3k)$, hence $d(k^2)/dm=\tfrac23 f'(m)$, which integrates to
$k^2-\tfrac23 f(m)=\mathrm{const}$. This relation coincides with
\eqref{eq:first-wavecurve} obtained for the first characteristic family
in Sec.~\ref{sec:whitham}. Its use in DSW fitting is nevertheless distinct from the construction of a global characteristic solution of the finite-amplitude Whitham system.

At the solitary-wave edge, the physical periodic
wavenumber vanishes. The boundary condition \(k(v_+)=0\) for the harmonic-edge fitting equation gives
\begin{equation}
k^2(m)=\frac23
\left[f(m)-f(v_+)\right].
\label{eq:harmonic_k_profile}
\end{equation}
Evaluating this expression at the harmonic edge,
where $m=v_-$, yields
\begin{equation}
k_h^2=\frac23
\left[f(v_-)-f(v_+)\right].
\label{eq:harmonic_edge_wavenumber}
\end{equation}

The harmonic-edge velocity is the linear group velocity evaluated at \((v_-,k_h)\), so that
\begin{equation}
s_h=f(v_-)-3k_h^2
=2f(v_+)-f(v_-).
\label{eq:harmonic_edge_speed}
\end{equation}

Equations \eqref{eq:harmonic_edge_wavenumber} and \eqref{eq:harmonic_edge_speed} constitute the harmonic-edge predictions of the DSW-fitting construction. Their derivation relies on the zero-amplitude dispersion relation and the associated simple-wave matching assumptions; it does not require an explicit finite-amplitude periodic-wave solution.

The harmonic-edge construction cannot be continued directly to the solitary-wave limit by using the fixed-wavenumber Stokes expansion. Indeed, the second-harmonic coefficient obtained in Sec.~\ref{sec:whitham}
contains the singular factor $k^{-2}$, and the weakly nonlinear expansion is not uniform as $k\to0$.

To determine the solitary-wave edge, we instead introduce the conjugate wavenumber $\widetilde{k}$. Analytic continuation of the linear dispersion
relation according to
\begin{equation}
\widetilde{\omega}(m,\widetilde{k})
=-i\omega_0(m,i\widetilde{k})
\label{eq:conjugate_definition}
\end{equation}
gives
\begin{equation}
\widetilde{\omega}(m,\widetilde{k})
=\widetilde{k}f(m)+\widetilde{k}^{\,3}.
\label{eq:conjugate_dispersion}
\end{equation}
Analogous harmonic-wave and soliton limits of the modulation equations, in which a linear wave or a soliton propagates on a dispersionless mean flow, have been analyzed for the Kadomtsev--Petviashvili--Whitham system~\cite{BiondiniHoeferMoro2020}.

The corresponding fitting equation is
\begin{equation}
\frac{d\widetilde{k}}{dm}
=
\frac{\partial_m\widetilde{\omega}}
{f(m)-\partial_{\widetilde{k}}
\widetilde{\omega}}
=-\frac{f'(m)}{3\widetilde{k}},
\label{eq:conjugate_fitting_ode}
\end{equation}
so that \(d(\widetilde{k}^{\,2})/dm=-\frac23 f'(m)\).

The conjugate wavenumber vanishes at the harmonic edge, where the physical periodic wavenumber remains finite. Imposing \(\widetilde{k}(v_-)=0\) and integrating gives
\begin{equation}
\widetilde{k}^{\,2}(m)
=\frac23\left[f(v_-)-f(m)\right].
\label{eq:conjugate_k_profile}
\end{equation}

At the solitary-wave edge,
\begin{equation}
\widetilde{k}_s^{\,2}
=\frac23\left[f(v_-)-f(v_+)\right].
\label{eq:soliton_conjugate_wavenumber}
\end{equation}

The solitary-wave edge speed is obtained from the conjugate phase velocity at \((v_+,\widetilde{k}_s)\), so that
\begin{equation}
s_s=f(v_+)+\widetilde{k}_s^{\,2}
=\frac23f(v_-)+\frac13f(v_+).
\label{eq:soliton_edge_speed}
\end{equation}

Equations \eqref{eq:harmonic_edge_speed} and
\eqref{eq:soliton_edge_speed} immediately imply
\begin{equation}
s_s-s_h=
\frac53\left[f(v_-)-f(v_+)\right]>0,
\label{eq:dsw_expansion_rate}
\end{equation}
since $f'(v)>0$ and $v_->v_+$.
The fitting construction therefore predicts an
expanding oscillatory region.

The edge ordering also satisfies the kinematic inequalities \(f(v_-)-s_h=2[f(v_-)-f(v_+)]>0\) and \(s_s-f(v_+)=\frac23[f(v_-)-f(v_+)]>0\).
Together with \eqref{eq:dsw_expansion_rate}, these are the
edge-speed inequalities expected for the adopted DSW
orientation. They are necessary kinematic consistency checks,
not sufficient conditions for a globally admissible
single-phase modulation solution.

The solitary-wave edge can be checked independently against
the traveling-wave reduction of the original dispersive PDE.
Setting \(v(\xi,\tau)=b+w(z)\) with \(z=(\xi-s\tau)/\varepsilon\) and \(b=v_+\), where \(w,w'\to0\) as \(|z|\to\infty\), and integrating the traveling-wave equation once using the far-field conditions gives
\begin{equation}
w''=(s-f(b))w-
\bigl[F(b+w)-F(b)-f(b)w\bigr].
\label{eq:solitary_profile_ode}
\end{equation}
A second integration yields the exact first integral
\begin{equation}
\frac12(w')^2=
\frac{s-f(b)}2w^2-
\int_0^w\bigl[F(b+y)-F(b)-f(b)y\bigr]\,\mathrm{d}y.
\label{eq:solitary_energy}
\end{equation}
For the quartic flux \eqref{eq:riemann_flux}, this becomes
\begin{equation}
(w')^2=w^2\left[s-f(b)-\frac{f'(b)}3w
-\frac{f''(b)}{12}w^2-\frac15w^3\right].
\label{eq:solitary_energy_polynomial}
\end{equation}
At the positive solitary-wave crest $w=A>0$, $w'=0$;
therefore the exact amplitude--speed relation is
\begin{equation}
s=f(b)+\frac{f'(b)}3A+\frac{f''(b)}{12}A^2
+\frac15A^3.
\label{eq:solitary_amplitude_speed}
\end{equation}
For the present nonlinearity, the derivative of the right-hand
side with respect to $A$ is
\begin{equation}
\frac{ds}{dA}=\frac{f'(b)}3+\frac{f''(b)}6A
+\frac35A^2>0\qquad(A\geq0).
\label{eq:solitary_speed_monotonicity}
\end{equation}
Indeed, its quadratic discriminant is
$[f''(b)]^2/36-4f'(b)/5<0$ for every real $b$,
while $f'(b)>0$. Consequently, for each $s>f(b)$
there is a unique positive crest amplitude $A$.
The bracket in \eqref{eq:solitary_energy_polynomial}
is strictly decreasing for $w\geq0$, is positive at $w=0$,
and vanishes at $w=A$. Hence the first integral supports
a positive homoclinic solitary-wave profile, up to translation.
This verifies the existence of a solitary-wave member with
the fitted speed, but does not prove that a global DSW
asymptotically matches that member.

The conjugate wavenumber gives an additional consistency
check. Linearization of \eqref{eq:solitary_profile_ode}
about $w=0$ yields the exponential decay rate
$\sqrt{s-f(b)}$ in the coordinate $z$. At the fitted
speed, $s_s-f(v_+)=\widetilde{k}_s^2$ by
\eqref{eq:soliton_edge_speed}, in agreement with
the conjugate-dispersion interpretation. Amplitude--speed relations of this type also govern the motion of gKdV solitons on slowly varying backgrounds~\cite{KamchatnovShaykin2023,vanDerSandeElHoefer2021}.

For the compressive Riemann data \eqref{eq:riemann_data},
the solitary-wave background is $b=v_+$.
Combining the exact amplitude--speed relation
\eqref{eq:solitary_amplitude_speed} with the
DSW-fitting velocity \eqref{eq:soliton_edge_speed}
gives
\begin{equation}
\frac{f'(v_+)}{3}A_s
+\frac{f''(v_+)}{12}A_s^2
+\frac15 A_s^3
=
\frac23\left[f(v_-)-f(v_+)\right].
\label{eq:solitary_dsw_amplitude}
\end{equation}
The strict monotonicity established in
\eqref{eq:solitary_speed_monotonicity}
ensures that this equation determines a unique
positive amplitude $A_s$ for every decreasing
jump $v_->v_+$.

Writing \(d=v_--v_+>0\) and expanding \eqref{eq:solitary_dsw_amplitude} as \(d\to0^+\) yields \(A_s=2d+O(d^2)\).
Thus the fitted solitary-wave amplitude recovers
the leading-order KdV-type small-jump scaling.

\subsection{Harmonic-edge hyperbolicity}
\label{subsec:edge-hyperbolicity}

The characteristic discriminant derived in Sec.~\ref{sec:whitham}
provides an additional diagnostic for the harmonic
edge. Evaluating it at the matching state
$(m,k)=(v_-,k_h)$ gives
\begin{equation}
\Delta_h
=
[f'(v_-)]^2
-3k_h^2f''(v_-).
\label{eq:delta_h_definition}
\end{equation}
Using \eqref{eq:harmonic_edge_wavenumber},
we obtain
\begin{equation}
\Delta_h
=
[f'(v_-)]^2
-2f''(v_-)
\left[f(v_-)-f(v_+)\right].
\label{eq:delta_h_explicit}
\end{equation}

The sign of $\Delta_h$ determines the leading-order
characteristic type predicted by the weakly nonlinear
Whitham system at the harmonic-edge matching state.
In particular, $\Delta_h>0$ corresponds to a real
splitting of the double zero-amplitude characteristic
branch, whereas $\Delta_h<0$ indicates a complex
splitting in the formal small-amplitude expansion.

For \(v_-\leq1/6\), one has \(f''(v_-)\leq0\), and consequently \(\Delta_h>0\).
For $v_->1/6$, the harmonic-edge discriminant
depends on the jump through $f(v_+)$.

A critical right state, when it exists in the
admissible interval $v_+<v_-$, is determined
implicitly by
\begin{equation}
f(v_+^{\mathrm{crit}})
=
f(v_-)
-\frac{[f'(v_-)]^2}{2f''(v_-)}.
\label{eq:critical_right_state}
\end{equation}
Because $f$ is strictly increasing, such a
critical state is unique whenever the right-hand
side lies in the range of $f$ over the selected
interval of right states.

For fixed \(v_->1/6\), the formal characteristic classification is \(\Delta_h>0\) for \(v_+>v_+^{\mathrm{crit}}\) and \(\Delta_h<0\) for \(v_+<v_+^{\mathrm{crit}}\).

The positivity of $k_h^2$ and the edge-speed ordering
in \eqref{eq:dsw_expansion_rate} hold for every decreasing
jump, independently of the sign of $\Delta_h$. Thus the
fitting kinematics and local weakly nonlinear hyperbolicity
are distinct requirements. In particular,
$\Delta_h>0$ is not, by itself, a sufficient global
DSW-admissibility criterion, and $\Delta_h<0$ does not
establish the nonexistence of every dispersive resolution.Indeed, DSWs have been found to persist across a hyperbolic-to-elliptic threshold of the dispersionless hydrodynamic equations in a quantum-droplet-bearing condensate, accompanied by additional wave patterns~\cite{ChandramouliEtAl2024}, and Gardner-type equations admit trigonometric bores, solibores, and kinks in place of classical DSWs~\cite{KamchatnovEtAl2012,KamchatnovEtAl2013}.

At $a=0$, the modulation matrix has a repeated eigenvalue,
so it is not strictly hyperbolic at the exact matching state.
The condition $\Delta_h>0$ instead ensures a real and
distinct splitting of that double branch for sufficiently
small positive amplitude, away from the degenerate threshold.

The equality $\Delta_h=0$ marks a degeneracy of
the leading-order characteristic splitting.
It should not, by itself, be interpreted as an
exact finite-amplitude modulational instability
threshold. Near this degeneracy, the ordinary
perturbation expansion of the characteristic
eigenvalues is nonuniform, and a separate local
analysis would be required to resolve the
transition.

To connect the preceding analysis with the numerical investigations, consider the representative Riemann data \(v_-=0.4\) and \(v_+=0.2\).
For these states, \(f(v_-)=0.368\) and \(f(v_+)=0.176\).
The harmonic-edge wavenumber is \(k_h=\sqrt{0.128}\simeq0.3578\),
and the DSW-fitting edge velocities are \(s_h=-0.016\) and \(s_s=0.304\).
For comparison, the dispersionless shock speed is \(s_{\mathrm{RH}}=4/15\simeq0.2667\).
Thus, \(s_h<s_{\mathrm{RH}}<s_s\).

The harmonic-edge discriminant takes the value \(\Delta_h=0.2704>0\).
The selected matching state satisfies the positive-discriminant
condition and therefore permits a real characteristic splitting
for sufficiently small positive amplitude. Strict hyperbolicity
does not hold at the exact zero-amplitude state.

For the fixed left state $v_-=0.4$, the critical
right state obtained from
\eqref{eq:critical_right_state} is approximately
\(v_+^{\mathrm{crit}}\simeq0.1422\).
The reference jump \(v_-=0.4\), \(v_+=0.2\) lies on the positive-discriminant side of this threshold.
For the same jump, the fitted solitary-wave speed \(s_s=0.304\) and background \(b=0.2\) give, from \eqref{eq:solitary_amplitude_speed}, \(A_s\simeq0.3947\), corresponding to a crest value \(v_{\mathrm{crest}}=v_++A_s\simeq0.5947\).
This is a traveling-wave consistency check, not an
independent measurement of the leading crest in the PDE
simulation.

These analytical values will serve as benchmarks
for the numerical study. In particular, the
solitary-wave edge speed provides a directly
measurable propagation characteristic, whereas
the harmonic-edge position requires a more
delicate interpretation of the small-amplitude
oscillatory tail.

The preceding construction combines two
complementary asymptotic descriptions. The
weakly nonlinear Whitham system characterizes
local periodic-wave modulation, including
hyperbolicity and genuine nonlinearity, while
DSW fitting determines the limiting edge
velocities through dispersion relations and
matching conditions.

These descriptions should not be regarded as
equivalent to an exact global solution of the
modulation problem.

In particular, the Stokes expansion is controlled by the relative magnitude of the second harmonic, measured by the indicator \(\chi(m,a,k)\) introduced in \eqref{eq:eta}.
The requirement \(\chi\ll1\) restricts the
finite-amplitude region in which the truncated
Whitham system can be used quantitatively.
The singular behavior of this indicator as
$k\to0$ prevents a uniform continuation of the
fixed-amplitude Stokes expansion to the
solitary-wave edge.

Furthermore, the characteristic integral curves
of the approximate system describe local
modulation directions, not an independently
established global self-similar DSW solution.
The existence of a characteristic wave curve emanating from a harmonic-edge state therefore does not, by itself, establish the complete matching to a solitary-wave edge.
Likewise,
the existence of the homoclinic traveling wave in
\eqref{eq:solitary_energy_polynomial} confirms a
compatible solitary-wave family but does not
supply the missing global modulation connection.

This distinction is especially relevant when
interpreting numerical harmonic-edge positions.
In an exact integrable Whitham description,
the edge may be identified through the
coalescence of appropriate modulation
invariants. In the present approximate system,
no corresponding exact global coalescence
condition has been established. Consequently,
a threshold-based harmonic-edge location
extracted from a direct numerical solution
need not coincide exactly with the formal
edge trajectory predicted by DSW fitting.
Such a discrepancy must be assessed together
with the numerical resolution, the edge
detection criterion, and the validity of
the underlying asymptotic assumptions.

The numerical investigation in the following section therefore distinguishes between three levels of information: the analytical DSW-fitting predictions, the local characteristic properties of the weakly nonlinear Whitham system, and the propagation features measured directly from solutions of the dispersive partial differential equation.
\section{Numerical Investigation of the Modulation Dynamics}
\label{sec:numerical}

The preceding analysis has established two complementary
descriptions of the dispersive dynamics. The weakly nonlinear
Whitham system provides the local characteristic structure of
modulated periodic waves, whereas the DSW-fitting construction
yields asymptotic predictions for the velocities of the
harmonic and solitary-wave edges. We now examine the
quantitative consequences of these results and assess their
consistency with direct numerical solutions of the modified
Gardner equation.

The numerical investigation addresses three related questions.
First, we determine whether characteristic continuation from
a prescribed harmonic-edge state encounters a loss of
hyperbolicity or genuine nonlinearity before the Stokes
expansion ceases to be asymptotically reliable. Second, we
examine how the harmonic-edge characteristic structure depends
on the Riemann data. Third, direct numerical simulations are used to assess the DSW-fitting predictions for the leading-edge velocity and, through the traveling-wave reduction, the amplitude and spatial profile of the leading solitary wave.

These comparisons are deliberately distinguished. Numerical
integration of the truncated characteristic equations tests
the internal behavior of the approximate modulation model,
whereas direct integration of the governing partial
differential equation provides an independent assessment of
the predicted DSW propagation.

\subsection{Characteristic continuation and Riemann-data dependence}
\label{subsec:numerical_characteristics}
\label{subsec:numerical_riemann}

We begin with the compressive Riemann data
\(v_-=0.4\), \(v_+=0.2\).
The harmonic-edge matching condition
\eqref{eq:harmonic_edge_wavenumber} gives
\(k_h=\sqrt{0.128}\simeq0.3578\), with corresponding
discriminant \(\Delta_h=0.2704>0\), so the initial state lies
in the strictly hyperbolic region of the leading-order
characteristic classification.

We integrate the \(\lambda_3\) wave-curve equations
\eqref{eq:wave-curves-pm} with \(\sigma=+1\), starting from
\(m(0)=v_-\) and \(k(0)=k_h\), and restricted to states with
\(\Delta>0\) and \(k>0\).
In addition to the discriminant, we monitor the truncated
genuine nonlinearity coefficient obtained from \eqref{eq:mu3},
denoted \(\mu_3^{\mathrm{tr}}\).

The applicability of the Stokes approximation is
monitored through the indicator \(\chi(m,a,k)\) introduced in
\eqref{eq:eta}.
The continuation is terminated when $\chi$ reaches
the prescribed value $\chi_*=0.2$, unless an earlier
loss of hyperbolicity or genuine nonlinearity occurs.
This threshold is a numerical consistency criterion;
it is not intended to define a sharp mathematical
boundary of validity.

Figure~\ref{fig:num_characteristic} shows the evolution
of $\Delta$, $\mu_3^{\mathrm{tr}}$, and $\chi$ along
the computed characteristic. The discriminant remains
strictly positive and increases along the trajectory.
The truncated genuine nonlinearity coefficient also
remains positive. Consequently, neither characteristic
coalescence nor a vanishing genuine nonlinearity
coefficient is detected before the prescribed
Stokes threshold is reached.

The terminal state is approximately
\(a_{\mathrm{end}}=0.1948\),
\(m_{\mathrm{end}}=0.3650\),
\(k_{\mathrm{end}}=0.2946\),
with
\(\Delta_{\mathrm{end}}=0.5238\) and
\(\mu_{3,\mathrm{end}}^{\mathrm{tr}}=1.5313\).

The significance of this calculation is not that
the truncated characteristic reaches the solitary-wave
edge. Rather, it identifies which of the competing
restrictions becomes active first along the selected
local continuation. For the present data, the
Stokes-consistency condition is encountered before
either of the characteristic degeneracies under
consideration. This observation is consistent with
the nonuniformity of the weakly nonlinear expansion
as the wavenumber approaches zero.

We next fix \(v_-=0.4\) and vary the right state
\(v_+\). From \eqref{eq:delta_h_explicit} one obtains
\(\Delta_h(v_+)=5.6f(v_+)-0.7152\).
Since \(f'(v)>0\), this function is strictly increasing; its
unique zero in the interval considered is the critical state
\eqref{eq:critical_right_state},
\(v_+^{\mathrm{crit}}\simeq0.1422\).

The condition $\Delta_h>0$ is satisfied for
$v_+>v_+^{\mathrm{crit}}$, while
$\Delta_h<0$ for $v_+<v_+^{\mathrm{crit}}$.
The latter region corresponds to a nonreal
characteristic splitting in the truncated
small-amplitude Whitham system. This classification concerns the approximate modulation equations
and should not be interpreted as a loss of
convexity of the underlying dispersionless flux $F$.

\begin{figure}[htbp]
\centering
\includegraphics[width=\linewidth]{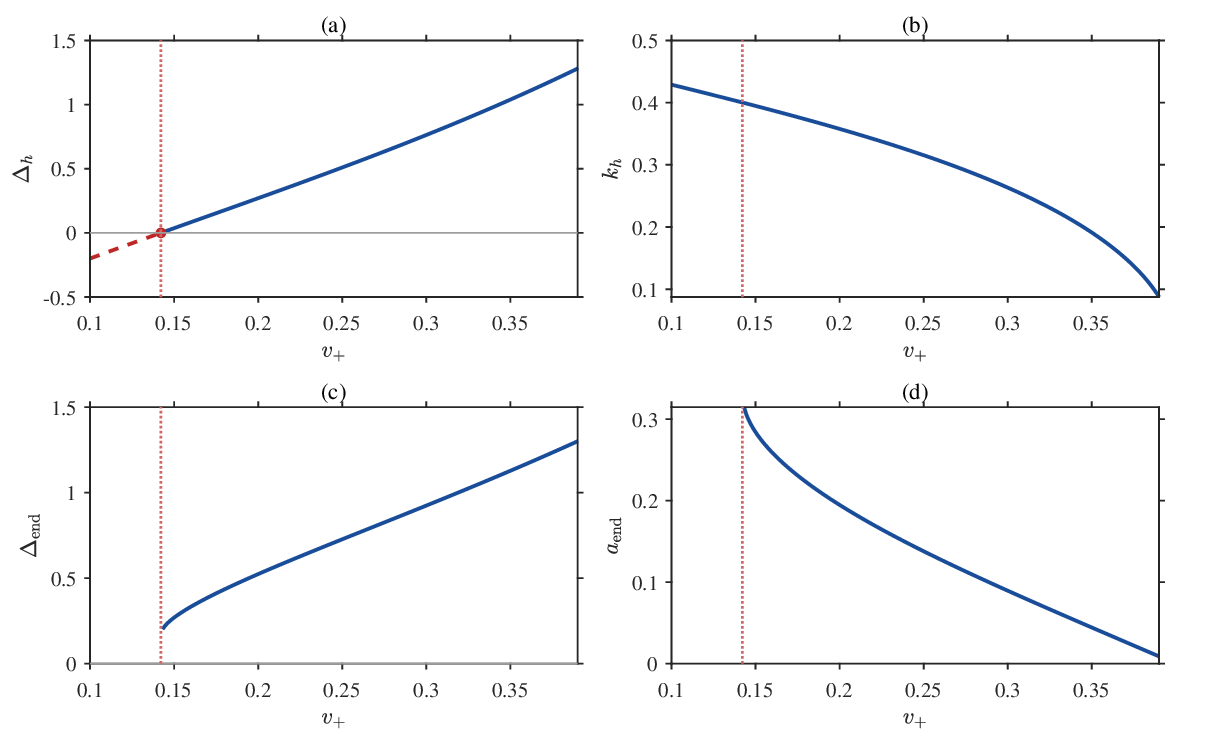}
\caption{
Dependence of the harmonic-edge and characteristic
continuation diagnostics on the right Riemann state
$v_+$, with $v_-=0.4$.
(a) Harmonic-edge discriminant $\Delta_h$.
(b) Harmonic-edge wavenumber $k_h$.
(c) Terminal discriminant $\Delta_{\mathrm{end}}$.
(d) Terminal amplitude $a_{\mathrm{end}}$.
The dotted vertical lines mark
$v_+^{\mathrm{crit}}\simeq\textcolor{blue}{0.1422}$.
The terminal quantities in panels (c) and (d)
are obtained only for initial states with
$\Delta_h>0$ and correspond to the stopping criterion
$\chi=0.2$.
}
\label{fig:num_riemann}
\end{figure}

Figure~\ref{fig:num_riemann} displays the dependence
of the harmonic-edge discriminant and wavenumber
on $v_+$, together with the discriminant and
amplitude at the termination of the corresponding
$\lambda_3$ characteristic continuation.

The parameter study includes 180 zero-amplitude
matching states with positive discriminant $\Delta_h$. In every computed
continuation, the prescribed Stokes threshold
is reached without an earlier detected loss
of hyperbolicity or genuine nonlinearity.
The terminal discriminant remains positive,
while the terminal amplitude decreases as
the right state approaches the left state.

Representative values are collected in
Table~\ref{tab:num_riemann}. The harmonic-
and solitary-wave edge speeds are evaluated
from the DSW-fitting expressions
\eqref{eq:harmonic_edge_speed} and
\eqref{eq:soliton_edge_speed}, respectively.
They are included to connect the characteristic
classification with the predicted macroscopic
propagation of the oscillatory region.

Two observations follow from these results.
First, the positive-discriminant boundary restricts
the Riemann data for which the selected real
characteristic splitting can be initiated. It is not
a universal threshold for DSW formation.
Second, the range of amplitudes accessible
under the prescribed Stokes criterion varies
substantially across the admissible states.

These findings characterize the internal
structure and limitations of the truncated
modulation model. They do not establish
global matching between the harmonic and
solitary-wave edges.

\subsection{Direct simulations and leading-edge comparison}
\label{subsec:numerical_pde}
\label{subsec:numerical_leading}
\label{subsec:numerical_resolution}

We now turn to the full dispersive equation
\eqref{eq:mg-conservative}, with the reference states
\(v_-=0.4\) and \(v_+=0.2\).

The discontinuous Riemann data are approximated
by a smooth transition. A distant compensating
transition is introduced to construct periodic
initial data compatible with the Fourier
pseudospectral discretization. The numerical
solution is evolved by a fourth-order
exponential time-differencing Runge--Kutta
method (ETDRK4), with the linear dispersive
operator represented in Fourier space and the
nonlinear flux evaluated pseudospectrally \cite{Trefethen2000,CoxMatthews2002,KassamTrefethen2005}. The same pseudospectral ETDRK4 approach has been applied to Gardner-type equations in cylindrical and spherical geometries ~\cite{AslanovaAhmetolanDemirci2020,AslanovaDemirciAhmetolan2022}

The simulations are performed for
\(\varepsilon=0.020\), \(0.010\), and \(0.005\),
with $N=2^{15}$ spatial grid points for
the first two cases and $N=2^{16}$ for
the smallest dispersion parameter.
The increased resolution in the latter
case is motivated by the shorter oscillatory
scales and by the resolution sensitivity
observed in preliminary computations.

\begin{figure}[htbp]
\centering
\includegraphics[width=\linewidth]{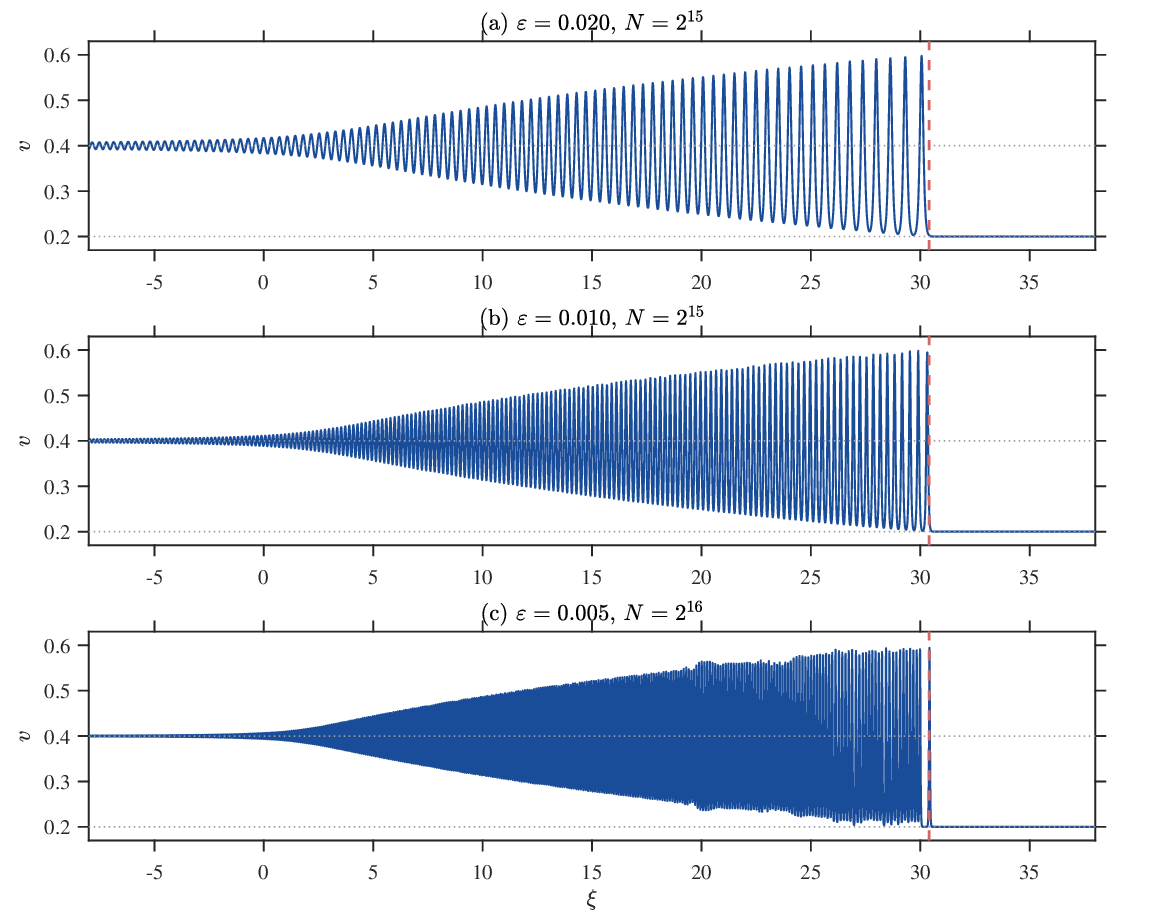}
\caption{
Direct numerical solutions of the modified
Gardner equation for smoothed Riemann data
with $v_-=0.4$ and $v_+=0.2$ at $\tau=100$.
(a) $\varepsilon=0.020$, $N=2^{15}$;
(b) $\varepsilon=0.010$, $N=2^{15}$;
(c) $\varepsilon=0.005$, $N=2^{16}$.
The vertical dashed line indicates the
DSW-fitting reference position
$\xi=s_s\tau$, with $s_s=0.304$.
The figure illustrates the increasing
oscillation density as $\varepsilon$
decreases and the approximate persistence
of the leading-edge propagation scale.
}
\label{fig:num_profiles}
\end{figure}

Figure~\ref{fig:num_profiles} shows the
numerical profiles at $\tau=100$.
The solutions exhibit an extended
oscillatory region whose local wavelength
decreases with the dispersion parameter.
At the same time, the macroscopic position
of the leading oscillatory structure
remains close to the DSW-fitting prediction.

The profile comparison is primarily qualitative:
agreement of the leading-edge positions does not imply
pointwise convergence of the rapidly oscillating solutions.
Small phase discrepancies can produce substantial pointwise
differences even when the large-scale envelopes remain comparable.

For the reference jump, the solitary-wave
edge velocity predicted by
\eqref{eq:soliton_edge_speed} is
\(s_s^{\mathrm{th}}=0.304\).
To estimate the corresponding velocity from direct
simulations, we track the position \(\xi_{\mathrm{peak}}(\tau)\)
of the leading oscillatory maximum.
Over \(50\leq\tau\leq100\), the trajectory is approximated by
the least-squares fit
\(\xi_{\mathrm{peak}}(\tau)=s_s^{\mathrm{num}}\tau+b\).

\begin{figure}[htbp]
\centering
\includegraphics[width=\linewidth]{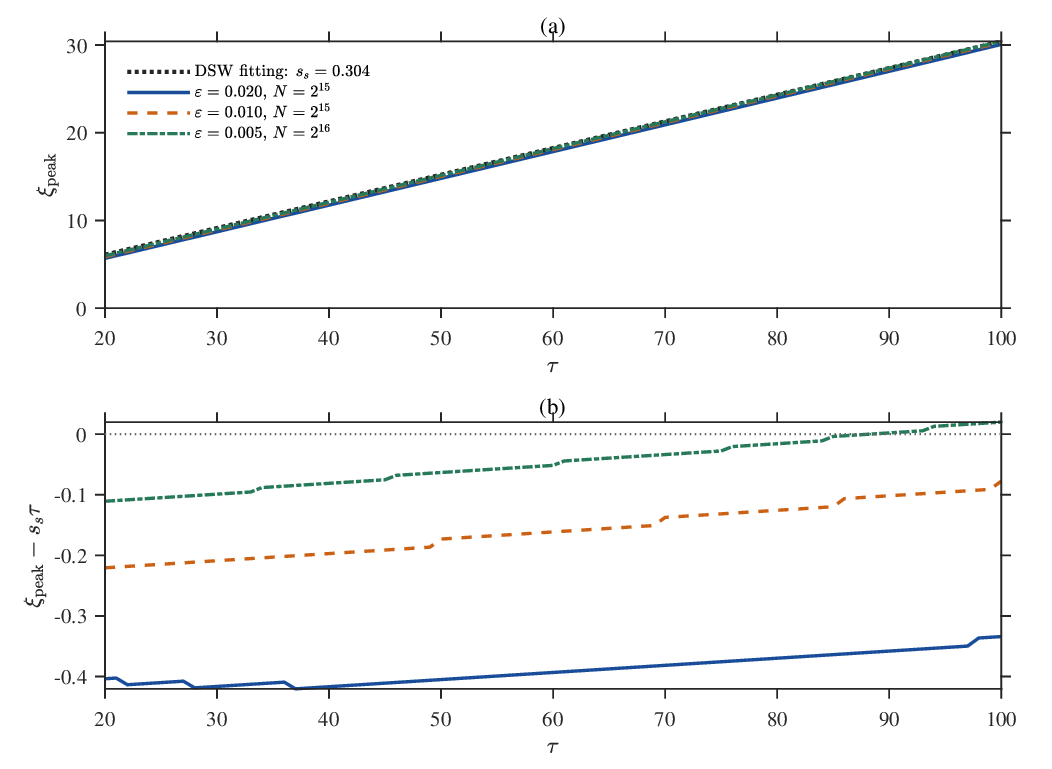}
\caption{
Leading-edge propagation for the Riemann
states $v_-=0.4$ and $v_+=0.2$.
(a) Numerically tracked leading-maximum
trajectories for three dispersion parameters,
compared with the theoretical DSW-fitting
reference $\xi=0.304\tau$.
(b) Position residuals
$\xi_{\mathrm{peak}}-0.304\tau$.
The measured propagation velocities
are obtained by least-squares fitting
over $50\leq\tau\leq100$.
}
\label{fig:num_edge_motion}
\end{figure}

Figure~\ref{fig:num_edge_motion}(a)
compares the resulting trajectories
with the theoretical reference line
$\xi=s_s^{\mathrm{th}}\tau$.
The corresponding position residuals
\(R(\tau)=\xi_{\mathrm{peak}}(\tau)-s_s^{\mathrm{th}}\tau\)
are shown in Fig.~\ref{fig:num_edge_motion}(b).

The fitted propagation speeds and
their relative deviations from the
theoretical prediction are reported
in Table~\ref{tab:num_edge_speeds}.

The measured velocities differ from
the theoretical prediction by less
than one percent in all three cases.
Moreover, their variation across
the selected dispersion parameters
is small compared with the magnitude
of the propagation speed.

This agreement supports the DSW-fitting prediction
for the leading-edge velocity. Velocity agreement must nevertheless
be distinguished from agreement of the absolute edge positions:
the residual decomposes as
\(R(\tau)=(s_s^{\mathrm{num}}-s_s^{\mathrm{th}})\tau+b\),
and therefore contains both a finite positional offset and a
contribution from the speed difference.

The near-linear late-time trajectories
indicate approximately constant
propagation speeds over the fitting
interval. However, the relative errors
do not decrease monotonically as
$\varepsilon$ is reduced. Consequently,
the present data do not determine an
asymptotic convergence rate in
$\varepsilon$.

We next comment on resolution sensitivity.
The accuracy of a Fourier pseudospectral
simulation of a DSW depends on the
resolution of its shortest dynamically
significant wavelengths. This requirement
becomes increasingly restrictive as
$\varepsilon$ decreases.

A separate resolution comparison was
performed for $\varepsilon=0.005$.
At $N=2^{15}$ and
$\Delta\tau=0.005$, the measured
leading-maximum speed was approximately
$0.3771$, whereas the calculation
with $N=2^{16}$ and
$\Delta\tau=0.0025$ yielded
$0.3057$.

The corresponding relative errors
with respect to the DSW-fitting
prediction were approximately
$24.0\%$ and $0.565\%$, respectively.
The lower-resolution solution also
exhibited pronounced irregularities
in the oscillatory region.

The comparison demonstrates that
the leading-edge measurement can be
sensitive to numerical resolution.
Because the spatial and temporal
resolutions were refined simultaneously,
these experiments do not separately
quantify spatial and temporal
discretization errors. They also do
not constitute a complete convergence
study.

The relative drift of the quadratic
invariant remained small in both
calculations, with maximum values of
approximately $1.91\times10^{-6}$
and $7.23\times10^{-7}$, respectively.
These diagnostics are useful but
insufficient on their own to establish
the accuracy of the resolved oscillatory
profile. In particular, the substantial
difference between the two numerical
profiles illustrates that near
conservation of an integral quantity
does not guarantee pointwise accuracy.

Accordingly, the high-resolution
$\varepsilon=0.005$ result is retained
for the leading-edge comparison,
while the lower-resolution calculation
is treated as a numerical sensitivity
test rather than as a physical
modulation regime.

Finally, the harmonic-edge velocity
predicted by
\eqref{eq:harmonic_edge_speed}
has not been independently validated
to the same degree as the solitary-wave
edge velocity. Near the harmonic edge,
the oscillation amplitude approaches
zero continuously, making a numerical
boundary location dependent on the
detection criterion and available
resolution. More importantly, the
truncated weakly nonlinear Whitham
system does not provide an exact
global self-similar solution whose
harmonic boundary can be located
through the coalescence of modulation
invariants. The present results
therefore do not establish quantitative
agreement for the harmonic-edge
trajectory.

\subsection{Solitary-wave amplitude and profile validation}
\label{subsec:numerical_solitary_validation}

The preceding comparisons establish the accuracy of the
DSW-fitting prediction for the propagation velocity of the
leading oscillatory structure. A more stringent assessment
is obtained by examining whether the same construction also
predicts the amplitude and spatial profile of the leading
solitary wave. This provides a complementary validation,
since the solitary-wave amplitude and profile are determined
from the traveling-wave reduction of the original dispersive
equation rather than from a direct fit to the numerical
solution.

For a compressive jump \(v_->v_+\), the DSW-fitting velocity
\(s_s\) is combined with the exact amplitude--speed relation
\eqref{eq:solitary_amplitude_speed}. Equivalently, the
predicted positive amplitude \(A_s\) is the unique solution
of \eqref{eq:solitary_dsw_amplitude}. The corresponding profile
follows from \eqref{eq:solitary_profile_ode} subject to
\(w(0)=A_s\), \(w'(0)=0\), and \(w(z)\to0\) as \(|z|\to\infty\),
with no amplitude or width adjustment to the PDE solution.

We fix \(v_-=0.4\) and take
\(v_+\in\{0.15,\,0.20,\,0.25,\,0.30\}\).
The simulations use \(\varepsilon=0.01\) up to \(\tau=100\),
with a Fourier pseudospectral ETDRK4 scheme,
\(N=2^{16}\) points, and \(\Delta\tau=0.0025\).

The leading numerical amplitude is measured as
\(A_{\mathrm{PDE}}=v_{\mathrm{crest}}^{\mathrm{PDE}}-v_+\),
with relative error
\(E_A=100|A_{\mathrm{PDE}}-A_s|/A_s\).
Crest-aligned profiles \(v_{\mathrm{PDE}}\) and \(v_{\mathrm{sol}}\)
are compared through
\[
E_2=
100\frac{\|v_{\mathrm{PDE}}-v_{\mathrm{sol}}\|_2}
{\|v_{\mathrm{sol}}-v_+\|_2},
\qquad
E_\infty=
100\frac{\|v_{\mathrm{PDE}}-v_{\mathrm{sol}}\|_\infty}
{A_s},
\]
over a common crest-centered window.

\begin{figure}[htbp]
\centering
\includegraphics[width=\linewidth]{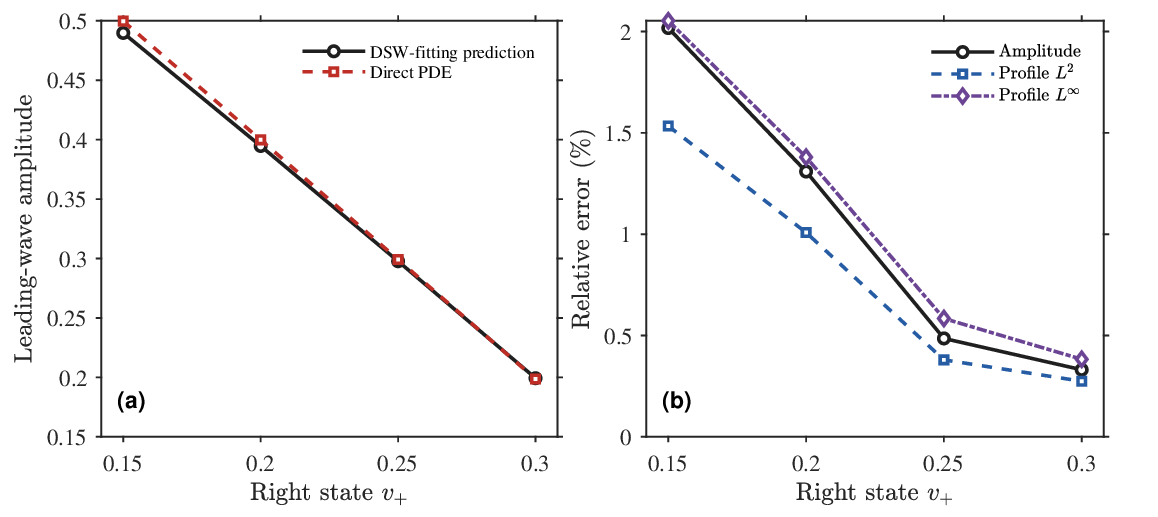}
\caption{
Validation of the leading solitary-wave predictions
for the fixed left state $v_-=0.4$ and four right
states $v_+$.
(a) Comparison of the theoretical amplitudes obtained
from DSW fitting and the traveling-wave amplitude--speed
relation with the amplitudes measured from direct PDE
simulations.
(b) Relative amplitude errors and crest-aligned
$L^2$ and $L^\infty$ profile errors.
All simulations are evaluated at $\tau=100$ with
$\varepsilon=0.01$.
}
\label{fig:num_amplitude_validation}
\end{figure}

The results are summarized in
Table~\ref{tab:num_solitary_validation} and
Fig.~\ref{fig:num_amplitude_validation}.
The predicted amplitudes decrease from approximately
$0.4898$ to $0.1993$ as the right state increases from
$0.15$ to $0.30$. The corresponding numerical amplitudes
follow the same dependence closely, with relative errors
ranging from $0.33\%$ to $2.02\%$.
The agreement is not restricted to the crest amplitudes:
the relative profile errors remain below $1.54\%$
in the $L^2$ measure and below $2.06\%$ in the
$L^\infty$ measure for all four cases.

The errors exhibit a systematic decrease across the
considered right states as the jump magnitude is reduced.
For the same sequence, the harmonic-edge discriminant
$\Delta_h$ increases from $0.0366$ to $0.7632$.
In particular, the case $v_+=0.15$ lies close to
the positive-discriminant boundary identified in
\eqref{eq:critical_right_state} and exhibits the
largest discrepancies. This correlation is consistent
with the increasing sensitivity expected near the
degenerate characteristic limit. Nevertheless, the
jump magnitude and $\Delta_h$ vary simultaneously,
so the present comparisons do not establish a
separate dependence of the approximation error on
the discriminant.

\begin{figure}[htbp]
\centering
\includegraphics[width=\linewidth]{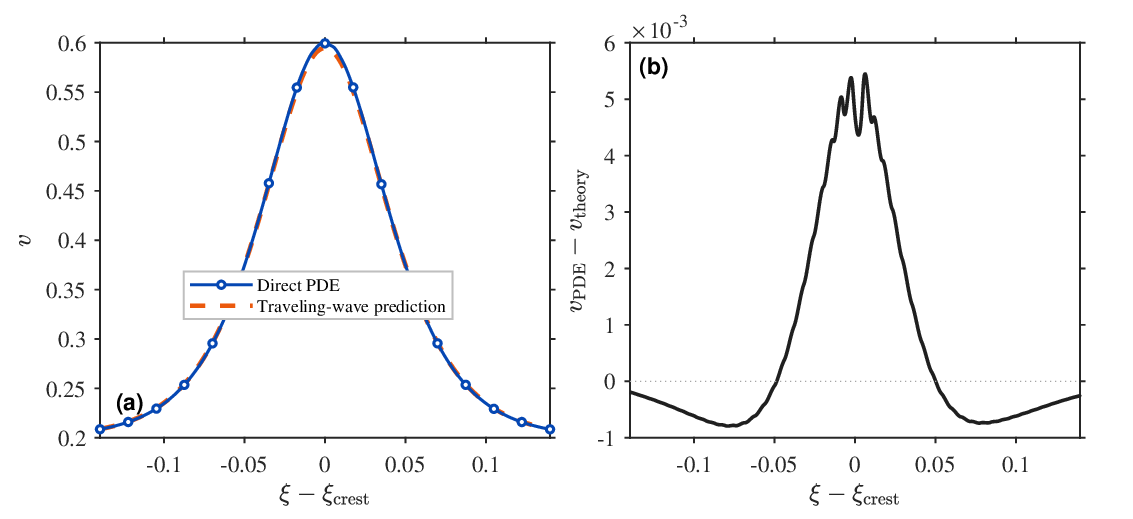}
\caption{
Leading solitary-wave profile validation for
$v_-=0.4$, $v_+=0.2$, $\varepsilon=0.01$,
and $\tau=100$.
(a) Crest-aligned profile extracted from the direct
PDE simulation and the solitary-wave profile predicted
by the traveling-wave equation using the DSW-fitting
velocity. No amplitude or width adjustment is applied.
(b) Pointwise difference between the numerical and
theoretical profiles in the crest-centered coordinate.
}
\label{fig:num_solitary_profile_validation}
\end{figure}

A more detailed comparison is shown in
Fig.~\ref{fig:num_solitary_profile_validation} for
the reference jump $v_-=0.4$, $v_+=0.2$.
The fitted velocity $s_s=0.304$ determines the
theoretical amplitude $A_s\simeq0.3947$ through
\eqref{eq:solitary_amplitude_speed}, corresponding
to a predicted crest value of approximately $0.5947$.
The numerical amplitude is approximately $0.3999$,
and the crest-aligned theoretical and numerical profiles
are nearly indistinguishable at the scale of the
full wave.

The pointwise difference in
Fig.~\ref{fig:num_solitary_profile_validation}(b)
is concentrated primarily near the wave crest and
remains of order $10^{-3}$ over the comparison window.
For this reference case, the relative profile errors
are approximately $1.01\%$ and $1.38\%$ in the
$L^2$ and $L^\infty$ measures, respectively.
The small residual oscillations near the crest may
reflect numerical resolution effects as well as
the fact that the leading wave is embedded in a
modulated dispersive wavetrain rather than being
an isolated solitary wave.

These comparisons provide a quantitative consistency
test of the combined DSW-fitting and traveling-wave
construction. The leading-edge velocity determines
the solitary-wave amplitude and profile, both of which
agree closely with the direct dispersive simulations
over the investigated family of Riemann data.
The crest alignment used for the profile comparisons
removes absolute positional differences; hence these
profile errors should not be interpreted as additional
measurements of the propagation velocity. The latter
is assessed independently in
Sec.~\ref{subsec:numerical_leading}.
Furthermore, the agreement does not establish a
globally valid finite-amplitude Whitham solution or
a rigorous matching theorem for the full DSW.

The combined analysis supports a limited conclusion.
For the selected compressive Riemann data, the positive-amplitude
characteristic continuation retains real distinct branches and a
nonvanishing truncated genuine nonlinearity coefficient up to the
imposed Stokes-consistency threshold \(\chi_*=0.2\), while direct
dispersive simulations yield leading-edge velocities close to the
DSW-fitting prediction. The independently predicted solitary-wave
amplitudes and crest-aligned profiles likewise agree quantitatively
with the PDE solutions across the investigated family of jumps.
The construction does not establish a uniformly valid finite-amplitude
Whitham solution throughout the oscillatory region.

\section{Conclusions}
\label{sec:conclusions}

We have investigated weakly nonlinear Whitham modulation and the dispersive resolution of compressive Riemann data for a modified Gardner equation with quadratic, cubic, and quartic nonlinear flux contributions. From a Stokes expansion and averaged conservation laws we obtained a three-component Whitham system and analyzed the perturbative splitting of its characteristic velocities, thereby distinguishing the existence of periodic traveling waves from the hyperbolicity and genuine nonlinearity of their slow modulations.

A principal result is the wavenumber-dependent change of characteristic type governed, to leading order, by the discriminant \eqref{eq:intro-discriminant}.
Because \(f'(m)>0\) for all real \(m\), the associated dispersionless flux \(F\) is strictly convex; the type change therefore does not arise from a loss of convexity of \(F\), but from the interplay between the curvature \(f''(m)\) and the carrier wavenumber.
For \(m\leq 1/6\), the leading-order splitting remains real for every positive wavenumber.
For \(m>1/6\), the curve \(k=k_c(m)\) separates a hyperbolic region, \(\Delta>0\), from an elliptic region, \(\Delta<0\), within the weakly nonlinear approximation.
The same discriminant at the harmonic-edge matching state yields a jump-dependent local diagnostic, which must not be conflated with a global criterion for DSW existence or admissibility: at zero amplitude the relevant eigenvalue is repeated, and the expansion is nonuniform near \(\Delta=0\).

For compressive Riemann data, DSW fitting yields explicit harmonic- and solitary-wave edge velocities without a global finite-amplitude Whitham solution.
The solitary-wave edge is further constrained by an exact traveling-wave energy relation.
Fourier pseudospectral simulations support the predicted leading-edge speed for the resolved cases considered here, with relative velocity discrepancies below one percent.
An independent comparison of leading solitary-wave amplitudes and crest-aligned profiles for four right states with \(v_-=0.4\) yields relative amplitude errors between approximately \(0.33\%\) and \(2.02\%\), crest-aligned relative \(L^2\) errors below \(1.54\%\), and relative \(L^\infty\) errors below \(2.06\%\).
Crest alignment does not independently validate the wave speed, and insufficient resolution can produce substantial edge-tracking errors.
The numerical agreement is therefore support for the leading-edge fitting prediction, not for a uniformly accurate Whitham description of the oscillatory interior.
In particular, a quantitative harmonic-edge trajectory and a global analytic DSW envelope have not been established.

Future work includes a distinguished-limit analysis near \(\Delta=0\), in the spirit of the modified nonlinear Schr\"odinger reduction derived at the instability boundary of the extended KdV equation~\cite{GrimshawEtAl2001} and of the hyperbolic--elliptic transition analysis of Refs.~\cite{BridgesRatliff2017,BridgesRatliff2019}, coupling the amplitude to the distance from the transition curve, to determine whether the apparent type change persists, shifts, or is regularized at higher order, and to clarify its relation to long-wavelength spectral stability \cite{JohnsonZumbrunBronski2010,JohnsonZumbrun2010,JinLiaoLin2019} without assuming that the two boundaries coincide at finite amplitude.
A second direction is the construction and validation of the full finite-amplitude Whitham system, which for gKdV-type equations can be expressed through period integrals of the traveling-wave potential~\cite{Whitham1974,JohnsonZumbrunBronski2010}, including its harmonic and solitary-wave degeneracies, to test whether both DSW edges can be joined by an admissible self-similar modulation.
Finally, Riemann data near the critical harmonic-edge discriminant, together with independent spatial and temporal convergence tests, would clarify how changes in local modulation type affect observable wave patterns. These problems extend the present edge-fitting analysis rather than replace it.
\section*{Decleration of competing interest}
The authors declare that they have no known competing financial interests or personal relationships that could have appeared to influence the work reported in this paper.
\section*{Acknowledgements}
This study was supported under Grant No. TGA-2025-46559  by Istanbul Technical University Office of Scientific Research Projects (ITU BAPSIS). 
\section*{Data availability}
No data was used for the research described in the article.

\bibliography{modified_gardner_references}

\newpage
\section*{Tables}

\begin{table}[htbp]
\centering
\caption{
Harmonic-edge quantities, DSW-fitting edge
speeds, and terminal characteristic amplitudes
for selected right states $v_+$, with $v_-=0.4$.
The terminal amplitude corresponds to
$\chi=0.2$. The edge velocities are analytical
fitting predictions, rather than quantities
measured from direct PDE solutions.
}
\label{tab:num_riemann}
\small
\begin{tabular}{@{}cccccc@{}}
\toprule
$v_+$ & $k_h$ & $\Delta_h$ &
$s_h$ & $s_s$ & $a_{\mathrm{end}}$\\
\midrule
0.15 & 0.394757 & 0.0366 & -0.0995 & 0.290083 & 0.284161\\
0.20 & 0.357771 & 0.2704 & -0.0160 & 0.304000 & 0.194808\\
0.30 & 0.263312 & 0.7632 &  0.1600 & 0.333333 & 0.089512\\
\bottomrule
\end{tabular}
\end{table}

\begin{table}[htbp]
\centering
\caption{
Numerically measured leading-edge
propagation speeds for $v_-=0.4$ and
$v_+=0.2$. The regression interval is
$50\leq\tau\leq100$ and the theoretical
reference speed is $s_s^{\mathrm{th}}=0.304$.
The percentage error is
$100|s_s^{\mathrm{num}}-s_s^{\mathrm{th}}|
/|s_s^{\mathrm{th}}|$.
}
\label{tab:num_edge_speeds}
\small
\begin{tabular}{@{}ccccc@{}}
\toprule
$\varepsilon$ & $N$ &
$s_s^{\mathrm{num}}$ & $b$ &
Error (\%)\\
\midrule
0.020 & $2^{15}$ & 0.30526 & -0.46911 & 0.413\\
0.010 & $2^{15}$ & 0.30584 & -0.27081 & 0.607\\
0.005 & $2^{16}$ & 0.30572 & -0.15210 & 0.565\\
\bottomrule
\end{tabular}
\end{table}

\begin{table}[htbp]
\centering
\caption{
Comparison of the DSW-fitting solitary-wave predictions
with direct PDE simulations for $v_-=0.4$,
$\varepsilon=0.01$, and $\tau=100$.
Here \(s_s\) is the theoretical leading-edge velocity,
\(A_s\) is obtained from \eqref{eq:solitary_dsw_amplitude},
and \(A_{\mathrm{PDE}}\), \(E_A\), \(E_2\), and \(E_\infty\)
are defined in the text. The quantity \(\Delta_h\) is the
harmonic-edge characteristic discriminant.
}
\label{tab:num_solitary_validation}
\small
\begin{tabular}{@{}cccccccc@{}}
\toprule
$v_+$ & $s_s$ & $A_s$ & $A_{\mathrm{PDE}}$
& $E_A$ (\%) & $E_2$ (\%)
& $E_\infty$ (\%) & $\Delta_h$\\
\midrule
0.15 & 0.290083 & 0.489808 & 0.499689
& 2.0175 & 1.5339 & 2.0531 & 0.03660\\
0.20 & 0.304000 & 0.394683 & 0.399851
& 1.3094 & 1.0082 & 1.3802 & 0.27040\\
0.25 & 0.318250 & 0.297722 & 0.299167
& 0.4855 & 0.3797 & 0.5839 & 0.50980\\
0.30 & 0.333333 & 0.199317 & 0.198658
& 0.3308 & 0.2743 & 0.3827 & 0.76320\\
\bottomrule
\end{tabular}
\end{table}

\clearpage
\section*{Figures}

For review purposes, all figures are collected after the references. They are
part of the main manuscript and are numbered consecutively in the order in
which they are cited in the text.

\begin{figure}[htbp]
    \centering
    \includegraphics[width=\linewidth]
    {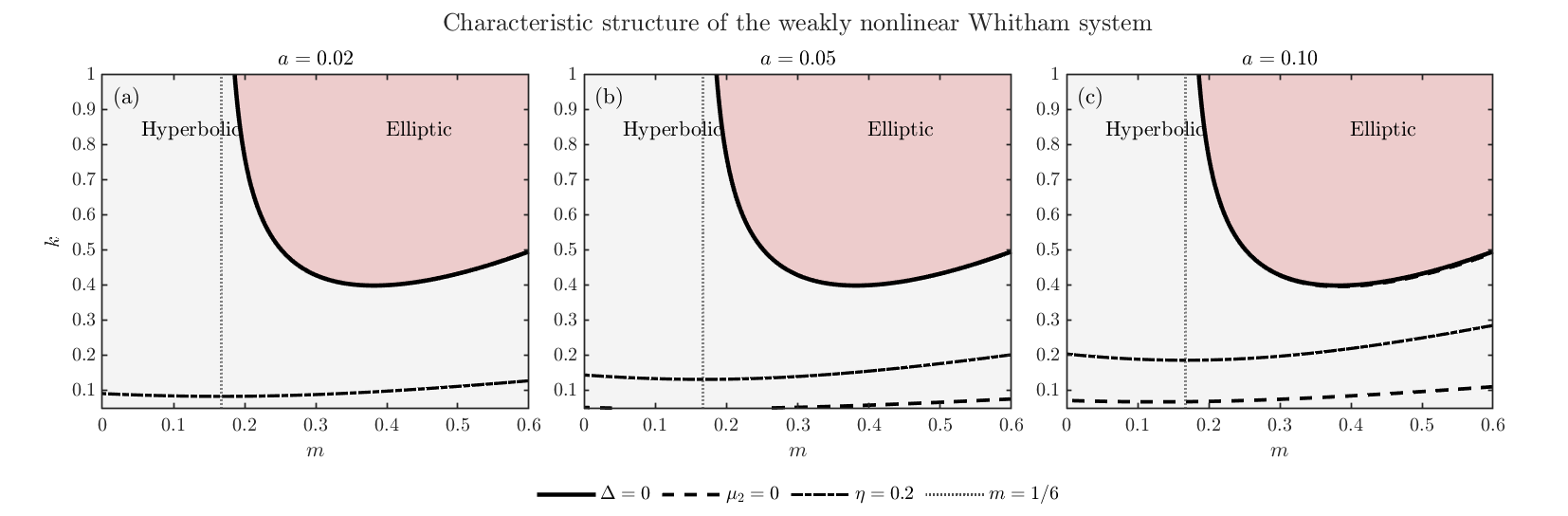}
    \caption{
    Characteristic structure of the weakly nonlinear Whitham
    modulation system for representative amplitudes
    $a=0.02$, $0.05$, and $0.10$.
    The solid curve $\Delta(m,k)=0$ separates the strictly hyperbolic
    region, $\Delta>0$, from the elliptic region, $\Delta<0$.
    The vertical dotted line denotes $m=1/6$, where $f''(m)=0$.
    The dashed curve represents the formal linear-degeneracy condition
    $\mu_2=0$, whereas the dash-dotted curve denotes the
    Stokes-consistency indicator $\chi=0.2$.
    The formal $\mu_2=0$ locus lies outside the conservative
    Stokes-consistency region indicated in the figure, and no
    $\mu_3=0$ locus occurs in the displayed hyperbolic domain.
    }
    \label{fig:characteristic-structure}
\end{figure}

\begin{figure}[htbp]
\centering
\includegraphics[width=\linewidth]{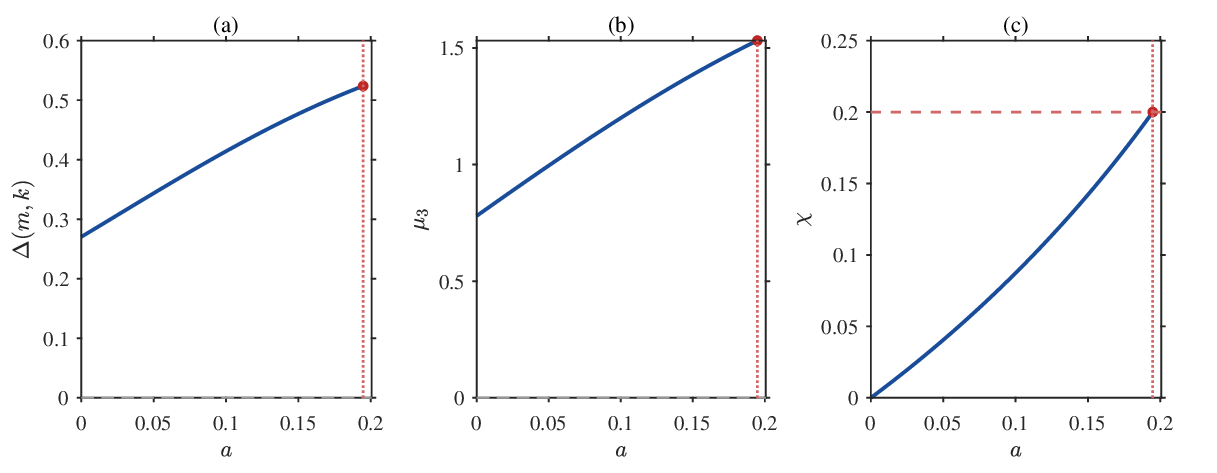}
\caption{
Continuation of the $\lambda_3$ characteristic
from the harmonic-edge state corresponding to
$v_-=0.4$ and $v_+=0.2$.
(a) Hyperbolicity discriminant $\Delta(m,k)$.
(b) Truncated genuine nonlinearity coefficient
$\mu_3^{\mathrm{tr}}$.
(c) Stokes-consistency indicator
$\chi=|af'(m)/(12k^2)|$.
The vertical dotted lines identify the terminal
amplitude at which $\chi=0.2$; the horizontal
dashed line in panel (c) marks this threshold.
No loss of hyperbolicity or genuine nonlinearity
is detected along the computed segment.
}
\label{fig:num_characteristic}
\end{figure}

\end{document}